\documentclass[11pt]{article}
\usepackage[margin=.75in]{geometry}
\usepackage[numbers,super]{natbib}

\usepackage{url}

\usepackage{graphicx}
\graphicspath{{figures/}}
\usepackage{multirow}

\usepackage{amsfonts}
\usepackage{bm}
\usepackage{bbm}
\usepackage{siunitx}

\usepackage{xspace}
\usepackage{relsize}

\usepackage{amsmath}

\def\E{\mathbb{E}}
\def\Prob{\mathbb{P}}
\def\Var{\mathbb{V}\mathrm{ar}}
\def\Cov{\mathbb{C}\mathrm{ov}}

\DeclareMathOperator*{\argmax}{argmax}

\def\PM{{PM}$_{2.5}$\xspace}

\usepackage{authblk}
\usepackage{fancyhdr}
\begin{document}
\renewcommand{\thefootnote}{\fnsymbol{footnote}}

\title{Case-crossover designs and overdispersion with application in air pollution epidemiology}

\lhead{Perreault et al.\ (2024)}
\rhead{Case-crossover designs and overdispersion}

\author[1,2]{Samuel Perreault\thanks{Corresponding author: samuel.perreault@utoronto.ca}}
\author[1,2,3]{Gracia Y. Dong}
\author[4]{Alex Stringer}
\author[5]{\\ Hwashin Shin}
\author[1,2]{Patrick Brown}

\affil[1]{{\small Department of Statistical Sciences, University of Toronto, ON, Canada}}
\affil[2]{{\small Centre for Global Health Research, St Michael’s Hospital, ON, Canada}}
\affil[3]{{\small Department of Mathematics and Statistics, University of Victoria, BC, Canada}}
\affil[4]{{\small Department of Statistics and Actuarial Science, University of Waterloo, ON, Canada}}
\affil[5]{{\small Environmental Health Science and Research Bureau, Health Canada, ON, Canada}}
	
\date{January 24, 2024}
	
\maketitle

\begin{abstract}
Over the last three decades, case-crossover designs have found many applications in health sciences, especially in air pollution epidemiology. 
They are typically used, in combination with partial likelihood techniques, to define a conditional logistic model for the responses, usually health outcomes, conditional on the exposures.
Despite the fact that conditional logistic models have been shown equivalent, in typical air pollution epidemiology setups, to specific instances of the well-known Poisson time series model, it is often claimed that they cannot allow for overdispersion.
This paper clarifies the relationship between case-crossover designs, the models that ensue from their use, and overdispersion.
In particular, we propose to relax the assumption of independence between individuals traditionally made in case-crossover analyses, in order to explicitly introduce overdispersion in the conditional logistic model.
As we show, the resulting overdispersed conditional logistic model coincides with the overdispersed, conditional Poisson model, in the sense that their likelihoods are simple re-expressions of one another.
We further provide the technical details of a Bayesian implementation of the proposed case-crossover model, which we use to demonstrate, by means of a large simulation study, that standard case-crossover models can lead to dramatically underestimated coverage probabilities, while the proposed models do not.
We also perform an illustrative analysis of the association between air pollution and morbidity in Toronto, Canada, which shows that the proposed models are more robust than standard ones to outliers such as those associated with public holidays.
\end{abstract}

\section{Introduction}\label{sec:intro}

In environmental epidemiology, a common problem is quantifying the association between daily counts of health events, such as deaths or hospitalizations, with short-term variations in environmental risk factors (exposures) such as temperature and air pollution. 
There are two commonly used methods to model these associations: Poisson time series models and conditional logistic models.\cite{Dominici/al:2003, Fung/al:2003, Ragettli/al:2017, Vicedo-Cabrera/al:2020}
The former is used to model the expected daily counts via a log-linear regression including, along with the exposures of interest, additional covariates to control for confounders; in particular, temporal covariates to capture the population's baseline hazard, which often exhibits long-term and seasonal trends.
In contrast, conditional logistic models make use of case-crossover designs\cite{Maclure:1991}, which compare each individual's exposure levels at the occurrence of the health event (the case day) with exposure levels on a set of days without a health event (control days).
Through a careful choice of the control days and the use of partial likelihood, case-crossover designs provide a natural way to control for the individuals' baseline hazard, and can thus greatly reduce the number of parameters that need estimation.

When exposures are shared across individuals, as we assume in this paper, the conditional logistic model can be seen a special case of Poisson time series.\cite{Levy/al:2001,Janes/Sheppard/Lumley:2005a,Lu/Zeger:2007}
This was established by showing that the estimating equations stemming from conditional logistic models indeed correspond to those of a Poisson time series model using a specific method for estimating the baseline hazard; this implicit baseline hazard model is determined by the underlying case-crossover design (the selection mechanism for control days).\cite{Lu/Zeger:2007}

Despite this equivalence, two reasons are often invoked for preferring Poisson time series models to case-crossover based conditional logistic models.\cite{Lu/Zeger:2007,Armstrong/Gasparrini/Tobias:2014}
The first is that fitting conditional logistic models give the appearance of being more computationally intensive, due to the need for an expanded dataset with additional 'control rows'.
However, given the equivalence between the two models, a possible solution is to perform inference using the augmented Poisson model corresponding to the specific case-crossover design, provided that this can be done efficiently.
Alternatively, Armstrong et al.\ (2014)\cite{Armstrong/Gasparrini/Tobias:2014} instead aggregate the data to daily counts and use a multinomial likelihood, which they refer to as conditional Poisson, in place of the conditional logistic.  
While some authors use the term "case-crossover" exclusively for models with conditional logistic likelihood, here we use \emph{case-crossover model} to refer to a model explicitly based on a case-crossover design, regardless of whether conditional logistic, multinomial (conditional Poisson), or an augmented Poisson likelihood is explicitly involved.

The second, and more consequential, concern proposed is that, in contrast to Poisson time series models, conditional logistic models cannot allow for overdispersion, i.e., to cases when the responses' variance exceeds the nominal (Poisson or logistic) variance\cite{McCullagh/Nelder:1989}.
Such belief seems to arise from the fact that overdispersion is usually introduced in Poisson time series models via quasi-likelihood functions, and that such methods are not applicable at the individual level. 
This is also the approach favoured by Armstrong et al.\ (2014)\cite{Armstrong/Gasparrini/Tobias:2014}, who, although they derive the conditional Poisson model from the conditional logistic, introduce overdispersion only after aggregation has been performed.

The purpose of this paper is two-fold.
The first objective is to clarify the relationship between case-crossover designs, case-crossover models and overdispersion.
As we argue, extra-Poisson variation in Poisson time series models has an intuitive meaning as uncaptured dependence among individuals.
This suggests that an appropriate time-to-event model at the individual level, upon which the conditional logistic model is based and from which the Poisson model can be derived via aggregation, should include some form of dependence between individuals.
We introduce such dependence at the individual level via shared daily random effects and, starting from the corresponding, more flexible version of the conditional logistic model, we derive an equivalent, aggregated, representation of the model akin to that of Armstrong et al.\ (2014)\cite{Armstrong/Gasparrini/Tobias:2014}, with the difference that overdispersion is now modelled explicitly at the individual level, rather than introduced after the aggregation step.

Our second objective is to highlight the importance of overdispersion in cases-crossover analyses.
To this end, we provide a Bayesian implementation of case-crossover models, which we use to perform an extensive simulation study and an illustrative analysis of environmental epidemiology data, where we measure the association between the concentration of fine particulate matters (\PM) and hospital admissions in Toronto, Canada.
Our simulation study provides strong evidence that standard (non-overdispersed) case-crossover models can lead to dramatically underestimated coverage probabilities, while their overdispersed analogues do not.
It also provides valuable insights into the bias-variance trade-off at play when the number of control days and/or the roughness of the baseline hazard is varied.
In the example, we focus on the overdispersion associated with public holidays, highlighting that the standard (non-overdispersed) case-crossover model is very sensitive to their inclusion or exclusion from the data while the overdispersed model is much more stable.

The paper is organized as follows.
In Section~\ref{sec:methods}, we begin by defining our time-to-event model via (dependent) subject-specific hazard functions, which we aggregate to obtain the corresponding population hazard function and the corresponding Poisson time series model.
We then discuss the conditional logistic and conditional Poisson models, with a focus on how the dependence introduced at the individual level gives rise to overdispersion.
We also briefly discuss the choice of reference frames and provide the technical details of our Bayesian implementation of the model.
The simulation study is presented in Section~\ref{sec:sim-study} and the example using real data in Section~\ref{sec:application}.
We conclude the paper with a short discussion of the main results and potential future works.
The Appendix, which contains several additional figures summarizing the results of the simulation study, and the scripts used to perform the simulation study and the real data analysis are provided as online supplementary material.
Our implementation of the model is available, as an \texttt{R} package, on GitHub.\footnote{https://github.com/samperochkin/bayesEpi}

\section{Methodology} \label{sec:methods}

\subsection{Hazard functions} \label{sec:hazard}

Suppose that we observe $N$ individuals for a period of $T$ consecutive time points.
For simplicity we assume data are observed on a
daily basis     throughout the paper.
For an individual $i$ on day $t$, let $Y_{it} = 1$ if individual $i$ had the event of interest on day $t$ and $Y_{it} = 0$ otherwise, and denote by $\rho_{it}$ its associated baseline hazard.
This latter incorporates all of individual $i$'s demographic information and is generally assumed to change smoothly with time $t$.
We consider time-to-event models for $Y_{it}$ given by the hazard function
\begin{equation} \label{eq:predictor}
\lambda_{it} = \rho_{it} \exp\left[\bm{X}_t^\top \bm{\beta} + 
\sum_{j=1}^{J} \gamma_j\left( U_{tj} \right) + Z_t \right]\;, \qquad
{Z_t} \sim  \mathcal{N}(0,\sigma_0^2)\;, \quad 
\gamma_j(\cdot) \sim \mathcal{GP}[\sigma^2_j Q_j(\cdot)]\;,
\end{equation}
where $\mathcal{N}$ refers to the normal distribution, $\mathcal{GP}$ refers to a centered Gaussian process, and, for $j=1,\dots,J$, $Q_j(\cdot)$ is a known stationary covariance function such that $\Cov\{\gamma_j(u+h),\gamma_j(u)\} = \sigma_j^2 Q_j(\|h\|)$, $u, h \in \mathbb{R}$.
Covariates for day $t$ are split into two vectors, $\bm{X}_t$ and $\bm{U}_t := (U_{t1},\dots,U_{tJ})$.
Those included in $\bm{X}_t$ are included in the model via linear fixed effects, while those included in $\bm{U}_t$ are included via random effects, and more precisely non-linear "exposure-response" functions $\gamma_j(\cdot)$, themselves modelled as Gaussian processes.
Commonly used linear covariates include day-of-week indicators and seasonal effects modelled using sinusoidal functions.
Exposure variables such as temperature and air quality are typically introduced via non-linear effects, either through $\bm{X}_t$ via splines or through $\gamma_j(\cdot)$.
A common model for $\gamma_j(\cdot)$ is the second order random walk\cite{Chiogna/Gaetan:2002, Lindgren/Rue:2008} (RW(2)), which is closely related to the second derivative penalty used in spline smoothing \cite{Hastie/Tibshirani:1990}; this is the model we use Sections~\ref{sec:sim-study} and \ref{sec:application}.
We describe its implementation in Section~\ref{sec:aghq-implementation}, where we further introduce identifiability constraints.

The remaining term, $Z_t$, is an additional daily random effect common to all individuals, and is not often seen in time-to-event models but is implicit in most Poisson time series models in air pollution epidemiology.
It is meant to capture the effects of events that are not included in any of the covariates, but may cause dependence among subjects.
Prime example of such events are public holidays: individuals may be less likely to go to the hospital on a public holiday, regardless of their exposures.
As we explain just below and in Section~\ref{sec:case-crossover}, the dependence introduced by the $Z_t$'s produces overdispersion; we thus we call them overdispersion effects.

When exposures are shared across individuals, it is possible and computationally advantageous\cite{Armstrong/Gasparrini/Tobias:2014} to consider as responses the daily case counts $Y_{t} := \sum_{i} Y_{it}$ instead of the individual health indicators $Y_{it}$.
In traditional time series analysis, where it is often assumed that $\rho_{it}$ vary only by a proportionality constant ($\rho_{it} = C_i \rho_{0t}$ for some constant $C_i$ and function $\rho_{0t}$) and that $\lambda_{it}$ is small, the conditional distribution of $Y_{t}$ is usually approximated by a Poisson distribution, that is,  $(Y_{t}|\bm{X}_t, \bm{U}_t, Z_t) \sim \mathrm{Poi}(\lambda_t)$ with
\begin{align} \label{eq:hazard-agg}
\begin{split}
    \lambda_t &=  \sum_i \lambda_{it}
    =  \sum_i C_i \rho_{0t} 
    \exp(\eta_t)
    = \exp(\mu_t + \eta_t),\\
    \eta_t &:= \bm{X}_t^\top \bm{\beta} + 
\sum_{j=1}^{J} \gamma_j\left( U_{tj} \right) + Z_t\;, \quad \mu_t := \log\left(\sum_i C_i \right) + \log(\rho_{0t})\;.
\end{split}
\end{align}
The population's log baseline hazard at time $t$, $\mu_t$, can be thought of as a nuisance function, and is typically modelled using a non-linear effect (i.e., splines or an autoregressive process).
This considerably simplifies inference, as it requires modelling only a single baseline hazard function, that of the population, rather than each individual one.
Note that the daily case counts $Y_t$ are sufficient statistics for $\lambda_t$ under the proportionality assumption, so that no information is lost during the aggregation process.

The strength of overdispersion in Poisson models based on \eqref{eq:hazard-agg} is controlled by the parameter $\sigma^2_0 = \Var(Z_t)$ from \eqref{eq:predictor}.
To see how $\sigma^2_0$ inflates the Poisson variance, let $\tilde \eta_t =  \bm{X}_t^\top \bm{\beta} +  \sum_j \gamma_j ( U_{tj} )$, so that $\eta_t =  \tilde\eta_t + Z_t$, and note that $(Y_t | \tilde\eta_t, Z_t) \sim \text{Poi}[\exp(\tilde\eta_t)\exp(Z_t)]$.
When conditioning on $Z_t$, $(Y_t|\tilde\eta_t, Z_t)$ is Poisson distributed with $\Var(Y_t | \tilde\eta_t, Z_t) = \E(Y_t | \tilde\eta_t, Z_t)$, i.e., the Poisson assumption is verified.
On the other hand, when conditioning only on $\tilde\eta_t$, $Y_t | \tilde\eta_t$ follows a Poisson-lognormal distribution\cite{poissonlognormal} and
$\Var(Y_t | \tilde\eta_t) = \E(Y_t | \tilde\eta_t) [1 + \E(Y_t | \tilde\eta_t)\{\exp(\sigma^2_0) - 1)\}]$,
which resembles the variance function $\phi \E(Y_t)$, for some $\phi > 0$, associated with quasi-likelihoods for Poisson regression.\cite{McCullagh:1983}
We further note that while the inclusion of a large number of non-Gaussian latent variables can considerably complicate inference, the choice of a Gaussian distribution for $Z_t$ is not in theory a necessity: for example, letting $\exp(Z_t)$ follow a gamma distribution leads to the negative binomial model with a similar variance function.\cite{VerHoef/Boveng:2007}

\subsection{Case-crossover models} \label{sec:case-crossover}

As the name suggests, case-crossover designs combine ideas from case-control and crossover designs\cite{Maclure:1991}: the main idea is to let each case serves as its own control.\cite{Maclure/Mittleman:2000}
For each individual $i$, exposures observed on the case day $t_i$ are compared with exposures observed on a set of control days.
Here, we restrict ourselves to situations where these are selected based on $t_i$ exclusively, for simplicity of notation, but the underlying argument can be easily extended to accommodate more intricate stratification schemes.
For any case day $t$, we denote by $\mathcal{T}(t)$ the set of control days associated with case day $t$, including the case day itself.
The key assumption underlying case-crossover models is that
\begin{equation}\label{eq:cbr}
\rho_{ir} \approx \rho_{is} 
\; ;\ r,s \in \mathcal{T}(t_i)
\end{equation}
for all individuals $i$.
In other words, it is assumed that the baseline risks corresponding to the days within a given reference frame are all similar.

When \eqref{eq:cbr} holds, for any individual $i$ and days $r,s \in \mathcal{T}(t_i)$, the ratio $\lambda_{ir}/\lambda_{is}$ is free of the nuisance parameters $\rho_{ir}$ and $\rho_{is}$.
It is thus invariant to any time-constant covariates contained in $\lambda_{it}$ such as demographic information (e.g., sex).
If each reference frame is entirely contained within a short window of time, it is also reasonable to suppose that the variation of $\lambda_{ir}/\lambda_{is}$ with respect to covariates whose effect changes relatively slowly over time, such as age, is negligible.
These assumptions are used to construct an approximate partial likelihood by conditioning, for any individual $i$, on the case day $t_i$ belonging to the reference frame $\mathcal{T}(t_i)$.\cite{Cox:1975,Breslow/al:1978,Holford/White/Kelsey:1978}
More precisely, assuming the the probability of repeated events for any individual is negligible, we can approximate the conditional probability density for the $N$ event times with the conditional logistic model
\begin{equation} \label{eq:likelihoodLambda}
\pi\left(t_1 \ldots t_N |T_i \in \mathcal{T}(t_i), i = 1 \ldots N \right) = \prod_{i=1}^{N} \Prob\left(t_i | T_i \in \mathcal{T}(t_i)\right)  = \prod_{i=1}^{N} 
\Big( \lambda_{it_i}  \bigg/ {\sum_{s \in \mathcal{T}(t_i)} \lambda_{is}} \Big).
\end{equation}
where $T_i$ to denote the random variable for individual $i$'s event date.
When setting $\sigma_0^2 = 0$, or equivalently $Z_t = 0$ for all $t$, \eqref{eq:likelihoodLambda} corresponds to a standard, non-overdispersed case-crossover model.

To exploit the fact that exposures are shared, denote by $\mathcal{T}_1,\dots,\mathcal{T}_K$ the $K < T$ unique reference frames among $\mathcal{T}(1),\dots,\mathcal{T}(T)$, and by $\bm{N}_k = (N_{ks}: s \in \mathcal{T}_k)$ the vector of daily counts of events from individuals associated with strata $k$, that is, $N_{ks} = \sum_{i \in \mathcal{I}_k} Y_{is}$ where $\mathcal{I}_k = \{i : \mathcal{T}(t_i) = \mathcal{T}_k \}$.
Aggregating the individuals within each of the unique reference frames results in counts that follow a multinomial distribution:
\begin{equation} \label{eq:ccmodel}
 (\bm{N}_{k}|\bm{\Delta}_k)  \sim 
\text{Multinomial}( \bar{N}_{k}, \bm{\Delta}_k )\;, \qquad \bar{N}_k := \sum_{s \in \mathcal{T}_k} N_{ks}\;, \quad \bm{\Delta}_k := (\Delta_{kt} : t \in \mathcal{T}_k)\;,
\end{equation}
where $\bar{N}_k$ represents the sum of events within the reference frame and $\bm{\Delta}_k$ is a vector of probabilities $\Delta_{kt} := 1/\sum_{s \in \mathcal{T}_k} \exp(\eta_s - \eta_{t})$.
It should be noted, however, that in \eqref{eq:ccmodel} we ignore potential information about the construction of reference frames.
In symmetric bidirectional designs, for example, the case day is always centered within the reference frame.
Failing to account for such features (in this case non-localizability\cite{Janes/Sheppard/Lumley:2005b}) may introduce bias.
In Sections~\ref{sec:sim-study} and \ref{sec:application}, we use so-called time stratified designs (see Section~\ref{sec:referent-frame}), for which \eqref{eq:ccmodel} holds with $N_{ks} = Y_s$.

Similar to Section~\ref{sec:hazard} for $Y_t$, we can investigate how $\bm{Z} := (Z_t)_{t=1}^T$ introduces overdispersion in the distribution of the stratified counts $N_{ks}$.
To this end, we define $\bm{\tilde\Delta}_k$ analogously to $\bm{\Delta}_k$ with the exception that it excludes the overdispersion effects, that is, we let $\tilde\Delta_{kt} := 1/\sum_{s \in \mathcal{T}_k} \exp(\tilde\eta_s - \tilde\eta_{t})$, with $\tilde\eta_{t}$ as in Section~\ref{sec:hazard}.
Conditional on $\bm{\Delta}_k$, we have that $\Var(N_{ks} | \bm{\Delta}_k) = \bar N_k \Delta_{ks} (1-\Delta_{ks})$.
Without conditioning on $\bm{Z}$, however, the counts follow a multinomial logistic random effects model\cite{multinomlogitnormal,binomiallogitnormal}, that is, $\Var(N_{ks} | \bm{\tilde\Delta}_k) = \bar N_k \tilde\Delta_{ks} (1-\tilde\Delta_{ks}) [1 + \sigma_0^2 (\bar N_k -1)\tilde\Delta_{ks} (1-\tilde\Delta_{ks})]$.
This last equation gives some more insight into the claim that the case-crossover framework as defined by Maclure\cite{Maclure:1991} is not compatible with overdispersion.
Notably, and as pointed out by McCullagh and Nelder (1989)\cite{McCullagh/Nelder:1989}, overdispersion cannot occur if $\bar N_k = 1$, since the term involving $\sigma_0^2$ in $\Var(N_{ks} | \bm{\tilde\Delta})$ then disappears and the standard multinomial variance is obtained.

Overdispersion also cannot occur, more generally, under the assumption that individuals are independent conditional on the covariates.
This occurs when the $Z_t$'s in \eqref{eq:predictor} are replaced with individual-specific overdispersion effect ($Z_{it}$), which makes them unidentifiable.
Thus, we can see that, in non-trivial situations, the seeming incompatibility of conditional logistic models with overdispersion stems from the assumption of independence between subjects, and not from the conditional logistic model or case-crossover design themselves.
This incompatibility even carries to the aggregated level, and is resolved by introducing dependence among the individuals through shared, unobserved variables in the hazard function, either implicitly via a quasi-likelihood function\cite{Armstrong/Gasparrini/Tobias:2014}, or more explicitly as we do in \eqref{eq:predictor}.

As it is more computationally efficient to work with the daily counts than the individual event times $T_i$,
we switch from the representation of the likelihood in \eqref{eq:likelihoodLambda} to an equivalent one based on the daily case counts $Y_t$ given by
\begin{equation} \label{eq:likelihood}
\log \pi\left(\bm{N}|\bm{\eta}\right) = \log \pi\left(\bm{Y}|\bm{\eta}\right) = - \sum_{t=1}^T Y_t \log\bigg(  \sum_{s \in \mathcal{T}(t)} \exp( \eta_{s} - \eta_{t})\bigg)\;, \qquad \bm{\eta} := (\eta_t)_{t=1}^T\;,
\end{equation}
where $\pi(\cdot)$ is the probability density function.
Following Armstrong et al.\ (2014)\cite{Armstrong/Gasparrini/Tobias:2014}, we refer to this formulation of our model as a \emph{conditional Poisson model}, to emphasize its strong connection with the Poisson time series model.
This is the formulation that we explicitly use when implementing the case-crossover model.

\subsection{Choice of reference frames} \label{sec:referent-frame}

The choice of reference frames requires careful consideration, as to minimize any biases that may arise.\cite{Redelmeier/Tibshirani:1997, Janes/Sheppard/Lumley:2005b}
In environmental epidemiology literature, control days are almost universally set to be on the same day of the week as the event day, for some number of weeks before or after the event week, to control for the day-of-week effect.
Translated in the aggregated, time series framework, this actually corresponds to a distinct time trend for each day of the week, and in particular to baseline risks $\rho_{it}$ in \eqref{eq:predictor} and $\rho_{0t}$ in \eqref{eq:hazard-agg} potentially showing significant jumps from one day to another.
Interestingly, day-of-week effects are often (though not always) included in Poisson time series models via day-of-week indicators\cite{Katsouyanni/al:1996, Liu:2019, Huang/al:2022}, which further constrains the seven time trends to be identical up to a single intercept parameter.

Popular schemes for selecting control days include the unidirectional ($k$ weeks preceding the event), symmetric bidirectional ($k$ weeks before and after the event) and time-stratified designs. 
Time-stratified designs involve partitioning the time interval covering the study into disjoint reference frames, such as calendar months or $k$-week intervals.
Analytical and simulation study results for (non-overdispersed, fixed effects) models have shown that bidirectional designs have lower bias than unidirectional ones in typical air pollution epidemiology scenarios\cite{Lumley/Levy:2000, bykov2020bias}, and it is generally acknowledged that time stratified designs are most effective.\cite{Janes/Sheppard/Lumley:2005a,Janes/Sheppard/Lumley:2005b,Mittleman:2005}
We refer the reader to an article of Lu and Zeger (2007) \cite{Lu/Zeger:2007} for a particularly enlightening presentation of case-crossover designs and their relation with time-trend estimation in Poisson time series.

The appropriate number of control days used to construct the reference frames necessarily depends on the roughness of the time trend in case counts, or more precisely on that of the possibly many time trends implicitly specified through the choice of design.
Increasing the number of control days results in lower uncertainty in parameter estimates, but can increase bias as $\rho_{0t}$ may become more heterogeneous within the reference frames.
This bias-variance trade-off is investigated and displayed in the simulation study of Section~\ref{sec:E2}.

\subsection{Bayesian implementation} \label{sec:aghq-implementation}

For usage in the simulation study and the application of Sections~\ref{sec:sim-study} and \ref{sec:application}, respectively, we propose a Bayesian implementation of (potentially overdispersed) case-crossover models.
It uses the approximate Bayesian inference methodology for case-crossover models without overdispersion recently developed and then formalized and generalized to extended latent Gaussian models in a series of papers by Stringer~et al.~(2021, 2022).\cite{stringer_approximate_2021, stringer_fast_2022}
This methodology relies on adaptive Gauss-Hermite quadrature, and is implemented using the \texttt{R} packages \texttt{aghq}\cite{aghq:2021} and \texttt{TMB}\cite{Kristensen/al:2016}.

In contrast to the prominent approach of drawing samples from the posteriors of interest using MCMC, our non-iterative approach is fast and scales well with the dimension of the parameter space, which is especially necessary given that the number of overdispersion effects equals the (potentially large) number of time points in the study.
This computational algorithm has some similarities with integrated nested Laplace approximation,\cite{Rue/Martino/Chopin:2009, Rue/al:2017} with which we implement our benchmark Poisson time series model in Section~\ref{sec:sim-study}.

We begin the description of our implementation by defining the Gaussian random processes $\{\gamma_{j}(\cdot)\}_{j=1}^J$ appearing, sometimes implicitly, throughout \eqref{eq:hazard-agg}--\eqref{eq:likelihood}.
For each $j=1,\dots,J$, we treat $\gamma_{j}(\cdot)$ as a Gaussian random walk of order two on a set of $K_j$ bins constructed by partitioning the interval of $\mathbb{R}$ covering the observed values $U_{1j}, \dots, U_{Tj}$ into a collection of contiguous intervals of equal width.
Let $\bm{\gamma}_{j}$ be such a random walk:
\begin{equation} \label{eq:gamma-approx}
     \bm{\gamma}_{j} := (\gamma_{j1},\dots,\gamma_{jK_j}) \sim \mathcal{N}(\bm{0},  \exp(-\theta_j) \bm{Q}^{-1}_j)\;,
\end{equation}
where $\bm{Q}_j$ is the $K_j \times K_j$ positive definite matrix characterizing random walks of order two (see Section~2 of Lindgren and Rue (2008)\cite{Lindgren/Rue:2008}). 
Because $\gamma_{j}(\cdot)$ is identifiable only up to the addition of a linear function, we further constrain the model by setting, for some reference value $u_j^*$ (the midpoint of some bin, say the $k$th) and any value $U_{jt}$ (falling, say, in bin $s$), $\gamma_j(U_{jt}) := \gamma_{js} - \gamma_{jk}$.
In practice, this is done by removing the $k$th and $(k+1)$th rows/columns of $\bm{Q}_j$ and incorporating an additional fixed effect $\beta_{j}'$ on $U_{tj} - u_j^*$ in the model.

Now, let $\bm{\gamma}$ concatenate the $J$ random effects vectors (of length $K_1-2,\dots,K_J-2$) and suppose that $\bm{\beta}$ includes their $J$ associated fixed effects; $\bm{\gamma}$ and $\bm{\beta}$ have dimensions $K_{1} + \dots + K_{J} - 2J$ and $\text{dim}(\bm{X}_t) + J$, respectively.
Furthermore, let $\bm{W} = (\bm{\beta},\bm{\gamma},\bm{Z})$ be the vector concatenating all the Gaussian latent variables and $\bm{\theta} := (\theta_0,\dots,\theta_{J})$ be the vector of log-precision parameters ($\sigma_j^2 = \exp(-\theta_j)$ for each $j=0,\dots,J$), and so that $\pi(\bm{Y}|\bm{W},\bm{\theta})$ is the conditional Poisson likelihood $\pi\left(\bm{Y}|\bm{\eta}\right)$ given in \eqref{eq:likelihood}.
The posterior distributions of interest are then
\begin{equation}\label{eq:posteriors}
	\pi(\bm{W}|\bm{Y}) = \int \pi(\bm{W}|\bm{Y},\bm{\theta})\  \pi(\bm{\theta}|\bm{Y})\ \mathrm{d} \bm{\theta} \qquad \text{and} \qquad 
	\pi(\bm{\theta}|\bm{Y}) = \int \pi(\bm{\theta},\bm{W}|\bm{Y})\ \mathrm{d}\bm{W}\;,
\end{equation}
which are both are intractable, and inferences thus are based on approximate posteriors.
To this end, let
$$
\widehat{\bm{W}}_{\bm{\theta}} =
\argmax\nolimits_{\bm{W}}\pi(\bm{W},\bm{\theta},\bm{Y}) \qquad \text{and} \qquad \bm{H}_{\bm{\theta}} = -\{\partial^{2}/(\partial \bm{W} \partial \bm{W}^\top)\}\log\pi(\bm{W},\bm{\theta},\bm{Y})\;,
$$
and let $\widetilde{\pi}_{G}(\bm{W}|\bm{\theta},\bm{Y})$ be the density of a $\mathcal{N}(\widehat{\bm{W}}_{\bm{\theta}},\bm{H}_{\bm{\theta}}^{-1})$ random variable.
Also, let
$$
\widetilde{\pi}_{LA}(\bm{\theta},\bm{Y}) = \frac{\pi(\bm{W},\bm{\theta},\bm{Y})}{\widetilde{\pi}_{G}(\widehat{\bm{W}}_{\bm{\theta}}|\bm{\theta},\bm{Y})}\;, \quad  \widehat{\bm{\theta}} = \argmax\widetilde{\pi}_{LA}(\bm{\theta},\bm{Y})\;, \quad \mathcal{H} = -\frac{\partial^{2}}{\partial \bm{\theta} \partial \bm{\theta}^\top}\log\widetilde{\pi}_{LA}(\bm{\theta},\bm{Y}),\;
$$
and $\mathcal{H}^{-1} = \mathcal{L}\mathcal{L}^{T}$, where $\mathcal{L}$ is the lower Cholesky triangle.
Now, for $\mathcal{Q} \subset \mathbb{R}^{J+1}$, define the nodes and weights of a product Gauss-Hermite quadrature rule of order $k$ in $\text{dim}(\bm{\theta}) = J+1$ dimensions\cite{Bilodeau/al:2022} as $\omega_k:\mathcal{Q}\to\mathbb{R}^{+}$.
The approximate posteriors are given by
\begin{align} \label{eq:inference}
\begin{split}
    \widetilde{\pi}(\bm{W}|\bm{Y}) &= |\mathcal{L}|\sum_{\bm{z}\in\mathcal{Q}}\widetilde{\pi}_{G}(\bm{W}|\widehat{\bm{\theta}} + \mathcal{L}\bm{z},\bm{Y})\ \widetilde{\pi}(\widehat{\bm{\theta}} + \mathcal{L}\bm{z}|\bm{Y})\ \omega_k(\bm{z})\;,\\
    \widetilde{\pi}(\bm{\theta}|\bm{Y}) &= \frac{\widetilde{\pi}_{LA}(\bm{\theta},\bm{Y})}{|\mathcal{L}|\sum_{\bm{z}\in\mathcal{Q}}\widetilde{\pi}_{LA}(\widehat{\bm{\theta}} + \mathcal{L}\bm{z},\bm{Y})\ \omega_k(\bm{z})}\;,
\end{split}
\end{align}
and inferences about the components of $\bm{W}$ are made by drawing exact samples from the approximate posterior $\widetilde{\pi}(\bm{W}|\bm{Y})$; see Algorithm~1 of Stringer~et al.~(2022)\cite{stringer_fast_2022}.

To complete the model, it remains to specify prior distributions for $\bm{\beta}$ and $\bm{\theta}$, which appear implicitly in \eqref{eq:posteriors}--\eqref{eq:inference} via $\pi(\bm{W},\bm{\theta},\bm{Y}) = \pi(\bm{Y}|\bm{W},\bm{\theta})\pi(\bm{W}|\bm{\theta})\pi(\bm{\theta})$, where $\pi(\bm{W}|\bm{\theta}) = \pi(\bm{\beta})\pi(\bm{\gamma}|\theta_1,\dots,\theta_J)\pi(\bm{Z}|\theta_0)$. 
Throughout Sections~\ref{sec:sim-study} and \ref{sec:application}, we always use independent, flat Gaussian priors (mean zero and precision $0.01$) for the linear fixed effects $\bm{\beta}$, including those added as part of the RW(2) models for the random effects, and a log-gamma prior with shape and rate parameters given by $0.5$ and $10^{-7}$, respectively, for the log-precision $\theta_0$ of the overdispersion effects, when included.
For the remaining log-precision parameters $(\theta_j)_{j=1}^J$, we use the priors which are induced by exponential priors on the standard deviations $(\sigma_j)_{j=1}^J$.\cite{Simpson/al:2017}
The choice of hyperparameter values vary depending on the situation.

\section{Simulation study} \label{sec:sim-study}

We now present a simulation study that compares the performance of the standard (non-overdispersed) and proposed (overdispersed) case-crossover models based on time stratified designs.
The simulation study is divided into two main experiments: The first experiment (Section~\ref{sec:E1}) is a comparison of the methods under idealized circumstances, whereas the second (Section~\ref{sec:E2}) presents a more realistic scenario, reflecting the fact that the functional form of the exposure-response is typically unknown.
We begin by describing how we generate synthetic data for the experiments.

\subsection{Data generation} \label{sec:ss-data-generation}

The simulation study uses \PM data for Toronto, Canada between 1996 and 2012 and a data generating model loosely based on a Poisson time series model fit to all-cause mortality data studied previously\cite{Huang/al:2022}. For simplicity we have generated data for a single city, although the signal in daily mortality data is weak and most studies report results obtained from many different regions  (i.e., over 600 cities in Liu~et~al.\ (2019)\cite{Liu:2019}).  To create simulated datasets with a sufficiently strong signal we doubled the measured \PM (Toronto has fairly low levels of air pollution), and doubled the average intensity of events by increasing the baseline risk.

\begin{figure}[t!]
\centering
\includegraphics[scale=.865]{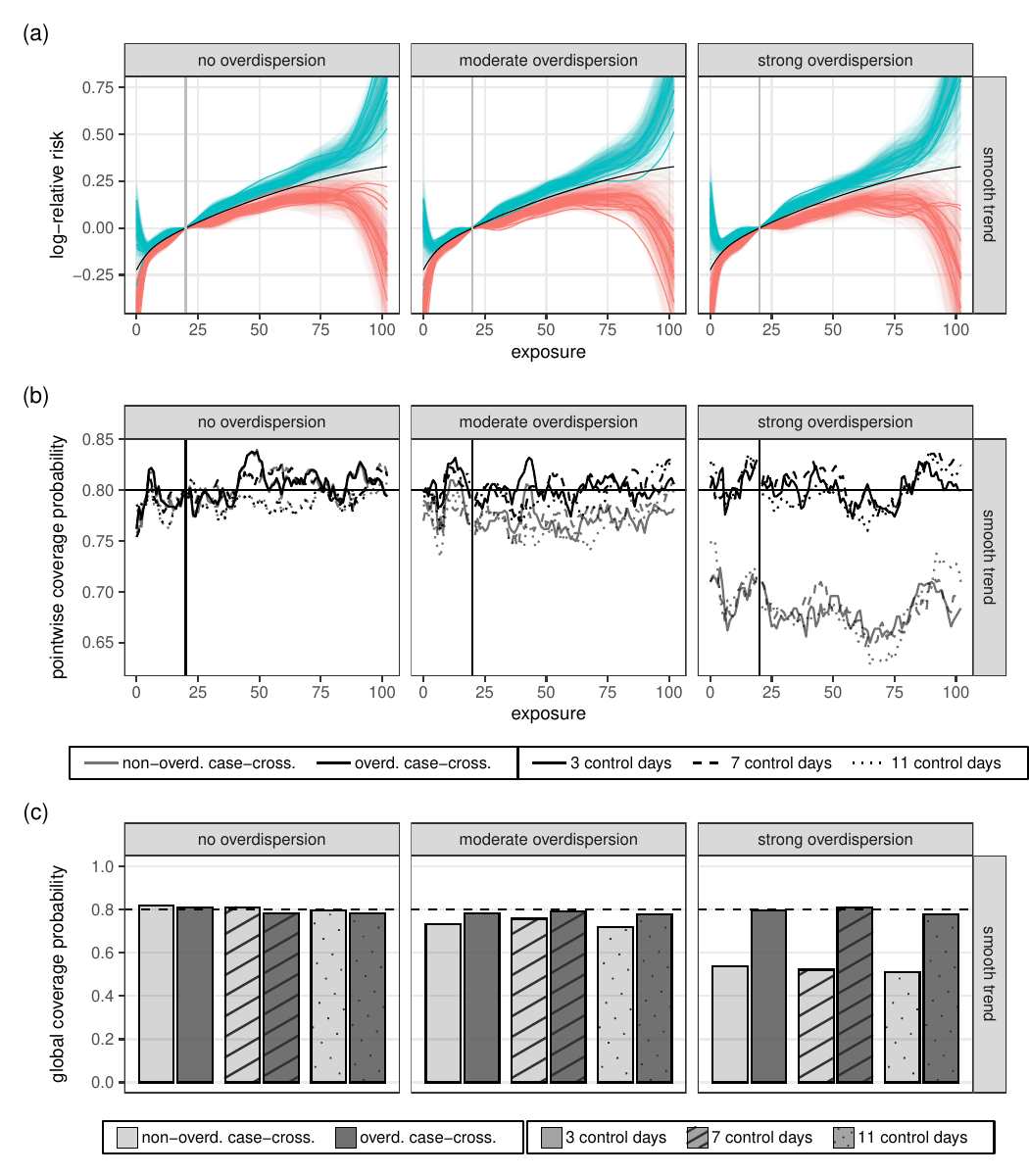}
\caption{Summary of the 500 simulation results from Experiment I (Section \ref{sec:E1}) for the smooth time trend and the three levels of overdispersion: none ($\sigma_0 = 0$), moderate ($\sigma_0 \approx 0.11$), strong ($\sigma_0 \approx 0.21$).
Six models were considered: non-overdipsersed and overdispersed case-crossover models based on time stratified designs with 3, 7 and 11 control days.
Panel (a) shows the estimated $80\%$ global envelopes (lower/upper limit in red/blue) returned by the overdispersed case-crossover model using 3 control days, with ten randomly selected envelopes shown more prominently;
the black curve depicts the true log-relative risk and the dark gray vertical line corresponds to the reference value 20.
Panel (b) shows the estimated coverage probabilities of the pointwise 80\% credible intervals; the horizontal and vertical lines correspond to the nominal level 80\% and the reference value 20, respectively.
Panel (c) shows the estimated coverage probabilities of the 80\% global envelopes; the dashed horizontal line corresponds to 80\%.
} \label{fig:E1}
\end{figure}

The concentration-response function was obtained by digitizing the reported exposure-response curve  for \PM on all-cause mortality reported by Liu~et~al.\ (2019)\cite{Liu:2019}.  A number of points on the curve were manually identified and a smooth curve constructed with regression splines.
More precisely, we used ten B-splines to fit the curves on each side of the reference value (setting it to \SI{20}{\micro\gram/\metre^3}), with respective knot vectors given by $(-0.2,-0.2,-0.2,-0.2,10,20)$ and $(20,30,50,102,102,102,102)$.  Using two separate sets of knot vectors ensures the curve is fixed at zero at the reference value, which simplifies fitting the case-crossover models in Experiment I (see Section~\ref{sec:E1-model}).
This concentration-response function was multiplied by 10, again in order to increase the signal in the simulated data.
The final exposure-effect curve is depicted in black in Figure~\ref{fig:E1}.a.

In addition to \PM, we included day-of-week indicators, seasonal terms (sinusoidal functions of $t$ with a one-year frequency), and a long-term trend as confounders in the generating models via fixed effects.
The fixed effects for the day-of-week indicators, from Sunday to Saturday, are given by $(0,0.2,0.3,0.3,0.25,0.2,0.05)$. 
Three different combinations of long-term and seasonal trends are considered: no trend, a smooth trend, and a rough trend; they are depicted in Appendix Figure~\ref{fig:time-trend}.
Note that in light of the relationship between Poisson and case-crossover models\cite{Lu/Zeger:2007}, the data generating model is roughly analogous to a case-crossover model based on a symmetric bidirectional design, with reference frames consisting of days immediately before and after the case day, and fixed effects for the day-of-week indicators.

Finally, for each combination of parameter values for the covariates, three distinct values for $\sigma_0$, the standard deviation of the overdispersion effects $Z_t$ in the generating model, are considered: $0$ (no overdispersion), $e^{-7/2}\approx 0.03$ (moderate overdispersion), and $e^{-3.5/2}\approx 0.17$ (strong overdispersion).  The moderate and high values are motivated by results obtained from the two real data examples in Section \ref{sec:application}.
The strength of overdispersion in the moderate and strong scenarios are roughly based on values estimated from the Toronto all-cause morbidity dataset (excluding and including public holidays, respectively) analyzed in Section~\ref{sec:application}.

\subsection{Experiment I: Fitting the true exposure-response model} \label{sec:E1}

\subsubsection{Models} \label{sec:E1-model}

For this experiment, we considered an idealized scenario where we fit the exposure-response curve as regression splines using the same basis functions (and knot locations), and reference value (\SI{20}{\micro\gram/\metre^3}) used when simulating the data.  Recall that the knots are unequally spaced, with more knots and hence greater curvature at low exposure values, and this analysis is unrealistically well adapted to estimating the true effect.  
We fit the case-crossover models of Section~\ref{sec:case-crossover} based on the time stratified reference scheme with 3, 7 and 11 control days.
We always used day-of-week as stratum variable and ensured that all reference frames consist of days from consecutive weeks, thus meaning that the partial likelihood in \eqref{eq:likelihood} is ``truly partial'', in the sense that we discarded some information; as discussed in Section~\ref{sec:referent-frame}, the fitted models implicitly use seven distinct time trends, while the time trend of the generating model is shared across all days of the week. 

\subsubsection{Results}
Figure~\ref{fig:E1} shows results simulated from the smooth time trend and the three different levels of overdispersion.
It shows the 80\% global envelopes returned by the overdispersed case-crossover using three control days (panel a), and the pointwise and joint coverage probabilities (panels b and c, respectively) for all models considered.
The joint coverage probabilities are based on global envelopes computed using the method of Myllymäki~et~al. (2017)\cite{Myllymaki/al:2017,Myllymaki/al:2020}.
See Appendix Figures~\ref{fig:E1-pc}--\ref{fig:E1-ci-od11} for a more comprehensive summary.
All results are based on 500 replications of each scenario considered.

The most evident feature is the below-expected coverage of the standard case-crossover model when the data are overdispersed, particularly in the rightmost panel (strong overdispersion) in Figure~\ref{fig:E1}.b.  The proposed overdispersed model consistently produces coverage near the expected level.
Additionally, since the standard case-crossover model is a special case of the overdispersed one, overdispersion can be safely included in case-crossover models even if there is no overdispersion in the data: When overdispersion is not present in the data, the two models yield very similar results.

Figure~\ref{fig:E1}.c further suggests, unsurprisingly, that selecting the width of the reference frame involves a bias variance tradeoff, as indicated by the slight decrease in coverage when the number of control days is increased.
This phenomenon is much more pronounced for scenarios involving a rougher time trend (see Appendix Figure~\ref{fig:E1-ge}).
Overall, the proposed overdispersed models show better results than standard case-crossover models.
Those using three control days, however, are the only ones seemingly unaffected by the strong time trend, while those using eleven control days perform the worst, producing noticeably underestimated coverage probabilities.
In general, for a given value of \PM, the model bias tends to increase and the width of the credible intervals tends to decrease with the number of control days used.

\begin{figure}[t!]
\centering
\includegraphics[scale=.865]{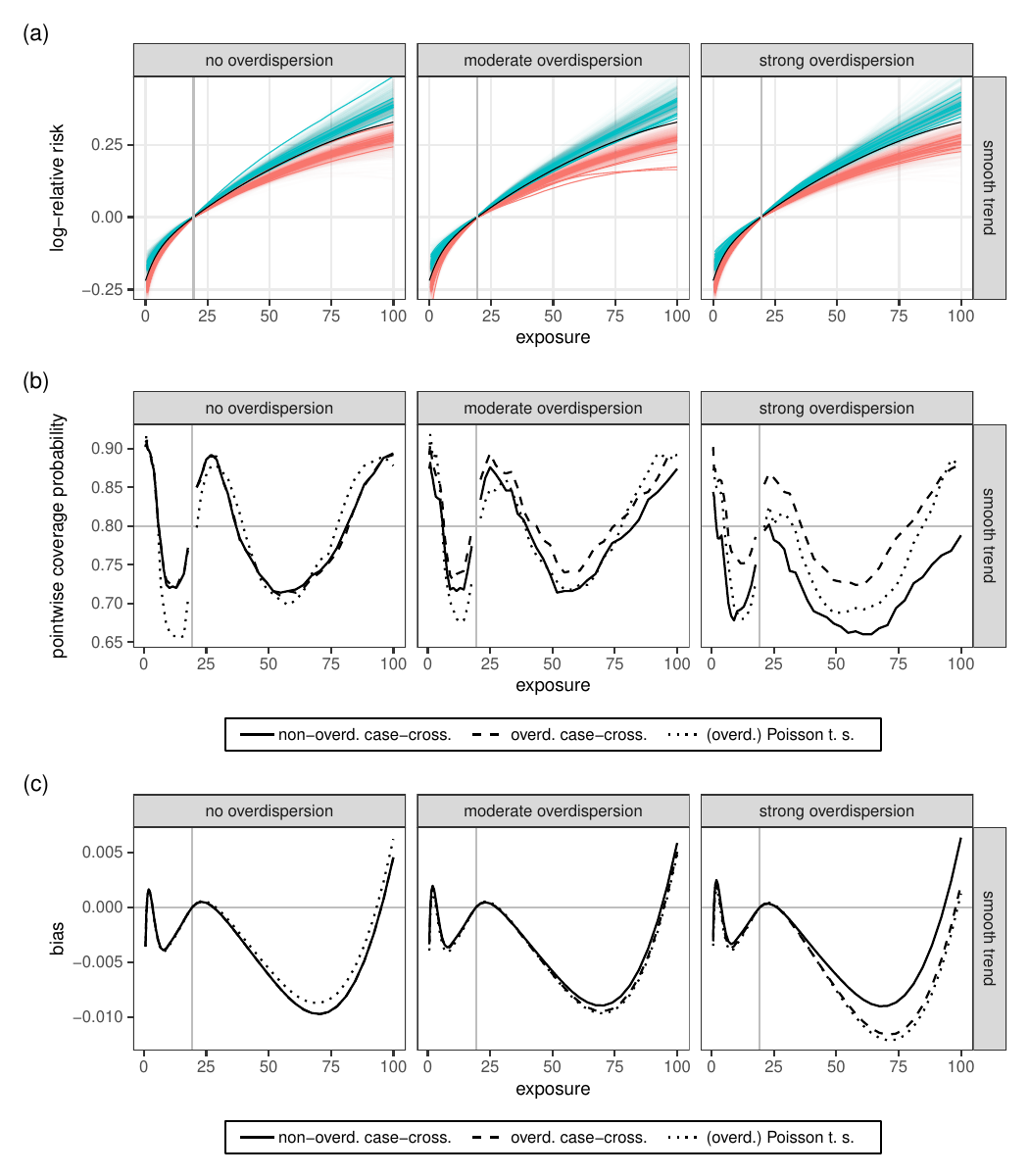}
\caption{Summary of the 500 simulation results from Experiment II (Section \ref{sec:E2}) for the smooth time trend and the three levels of overdispersion, for each of the three models considered: the non-overdispersed and overdispersed case-crossover models (time stratified with three control days, depicted by contiguous and dashed lines, respectively), and the Poisson time series model (depicted by dotted lines).
Panel (a) shows the estimated $80\%$ global envelopes (lower/upper limit in red/blue) returned by the overdispersed case-crossover model (time stratified, three control days), with ten randomly selected envelopes shown more prominently;
the black curve depicts the true log-relative risk and the dark gray vertical line corresponds to the reference value 20.
Panel (b) shows the estimated coverage probabilities of the pointwise 80\% credible intervals; the dark gray horizontal and vertical lines correspond to the nominal level 80\% and the reference value 20, respectively.
Panel (c) shows the estimated bias of the posterior median point estimates; the gray horizontal and vertical lines correspond to 0 (no bias) and the reference value 20, respectively.
} \label{fig:E2}
\end{figure}

\subsection{Experiment II: Gaussian process exposure-response model}\label{sec:E2}

\subsubsection{Models}

In contrast with Experiment I, where the smoothness of the exposure-response function was determined by pre-specified knot locations, here we used a flexible Gaussian process model where the roughness of the curve is a parameter to be inferred from the data.  More specifically, we modelled the exposure-response curve through $\gamma_1(\cdot)$ in \eqref{eq:hazard-agg} using a random walk of order two as described in Section~\ref{sec:aghq-implementation}, with $\sigma_1$ determining the smoothness of the curve.
We further provided some degree of subject-matter expert knowledge by introducing \PM in the model as $(\text{\PM})^{1/2}$, to naturally favour square-root-like shaped curves.
On this new scale, we used bins of width 0.2 when discretizing the range of values taken by $(\text{\PM})^{1/2}$, and we set the reference value to \SI{0.4}{\micro\gram/\metre^3}.

As prior for $\sigma_1$, we used an exponential distribution, which favours values close to zero and hence encourages the exposure effect to be a linear function of the exposure (or a square root function of \PM on the original scale).
A prior median of $4 \times 10^{-4}$ for $\sigma_1$ was used, a value that was found to have the expected global coverage for non-overdispersed data and the standard case-crossover model. Increasing the prior median lead to conservative results and much higher than expected coverage, which would not necessarily be undesirable in practice but for the purposes of a simulation study the stronger prior provided a clearer illustration of the methods.

We fit case-crossover models using the same time-stratified reference scheme as in the previous experiment, this time restricting our attention to three control days for convenience.
As benchmark, we also considered a Poisson time series model based on the data generating one, which we fit using the \texttt{INLA}\cite{Rue/al:2017} package in \texttt{R}. 

\subsubsection{Results} \label{sec:E2-results}
Figure~\ref{fig:E2} summarizes the results for the smooth trend and all three levels of overdispersion.
It shows the 80\% global envelopes returned by the overdispersed case-crossover model (panel a), also computed using the methodology of Myllymaki~et~al.~(2017)\cite{Myllymaki/al:2017,Myllymaki/al:2020}; the pointwise coverage probabilities for all three models considered (panel b); and the pointwise bias of the corresponding median effects (panel c).
See Appendix Figures~\ref{fig:E2-pc}, \ref{fig:E2-bias} and \ref{fig:E2-ci-nod}--\ref{fig:E2-ci-p} for the results specific to each trend.
Again, all results are based on 500 replications of each scenario considered.

As in Section \ref{sec:E1}, the standard case-crossover and the proposed overdispersed models yield results that are extremely similar when the data are not overdispersed, and the performance of the standard case-crossover model deteriorates with the inclusion of overdispersion in the data.
Overall, the overdispersed case-crossover and Poisson models give similar results, with pointwise credible intervals being reasonably accurate, especially for larger values of the variable of interest.
The most striking difference with the results of Section \ref{sec:E1}, apparent when comparing Figures~\ref{fig:E1}.a and \ref{fig:E2}.a, is that the resulting estimates are now much smoother, especially for \PM values at both extremes of their range.
In these regions, which contain few data points, the B-splines model has little information to work with; in contrast, the RW(2) model, which implicitly penalizes large second differences, favours linear interpolation.

This latter fact also explains, at least partially, the variation in pointwise coverage probabilities.
The RW(2) model assumes that the exposure-effect curve's second differences are independent, but this is not satisfied, as the true exposure-effect curve is not exactly a scaled square-root function.
Such discrepancy between the model and the true exposure-effect likely introduces the estimation bias depicted in Figure~\ref{fig:E2}.c and Appendix Figure~\ref{fig:E2-bias}.
Interestingly, adjusting the limits of the pointwise credible intervals by subtracting the bias, which is small (at most $\pm 0.1$) relative to the exposure-effect, leads to much more reasonable, conservative coverage probabilities (mostly between 85\% and 90\%, for an 80\% nominal level), as depicted in Appendix Figure~\ref{fig:E2-pc}.
This suggests that the original intervals produced by the overdispersed models are quite accurate after all.

The Poisson time series has produced results which appear superior to the case-crossover models, coverage probabilities are competitive and the credible intervals are smaller (see Appendix Figure~\ref{fig:E2-width}).  The time series model fit here imposes a stronger set of assumptions than the case crossover model: the trend is smoothly varying and day-of-week effects are additive and constant over time.  In this simulation study these assumptions are accurate and the additional structure assumed is leveraged by the model to reduce uncertainty without contributing to bias.  The case-crossover models have a much weaker set of assumptions, i.e.\ allowing for entirely different baseline hazards for each day of the week, which increases uncertainty in the estimates.  In practice the choice between a time series model and a case-crossover analysis is partly subjective, and rests on how confident one is in the former's assumptions.

\section{Application to air pollution epidemiology} \label{sec:application}

\subsection{Data and model} \label{sec:app-data}

As illustrative example, we now use the case-crossover model given in \eqref{eq:likelihood} to measure the association between daily non-accidental morbidity counts (hospitalizations having ICD-10 codes A00-R99) and ambient concentrations of particulate matters of diameter less than \SI{2.5}{\micro\metre} (\PM, in \SI{}{\micro\gram/\metre^3}) in Toronto, Canada.
Our analysis is based on data from Huang~et~al.~(2022)\cite{Huang/al:2022}, which covers the years 1996 to 2012 ($T=6210$ days), for the 36--70 age group.

We focus in particular on the contribution of holidays to overdispersion, and their effect on the fit of both the standard and overdispersed case-crossover models.
We included end-of-year holidays, spanning from the Wednesday in the week before Christmas to the first Thursday in the week after the New Year, Easter holidays, spanning from Good Friday to Easter Monday, Victoria day, Canada day, Civic Holiday, Labour Day, Thanksgiving Day, and Remembrance Day.
It is generally acknowledged that the effect of holidays should be checked when studying the association between air pollution and health outcomes.\cite{Schwartz/al:1996,Katsouyanni/al:1996}
As we demonstrate, this is easily done with our proposed method by analyzing the estimated overdispersion effects $(Z_t)_{t=1}^T$.
For our data, we find that holidays indeed correspond to outlying observations, and that the standard case-crossover design is sensitive to their inclusion or exclusion in the data, while its overdispersed analogue is more robust in this respect.

In addition to the health outcome and \PM, we consider temperature (in \SI{}{\celsius}) and relative humidity (in \%) as confounders in the model.
We introduce relative humidity via a linear fixed effect, and \PM (on the square root scale) and temperature via non-linear random effects; that is, in \eqref{eq:hazard-agg}, we have $X_t$ as the humidity on day $t$ and $\bm{U}_t$ is a vector of $(\text{\PM})^{1/2}$ and temperature.
We use exponential priors with medians $0.002$ and $0.005$ for the standard deviations $(\sigma_1,\sigma_2)$ associated with \PM (on the square root scale) and temperature, respectively.
Following earlier work\cite{Di/al:2017,Liu:2019,Huang/al:2022}, we model the health outcome on day $t$ using lagged averages of the exposures: two days lags (average exposure over days $t$ and $t-1$) for \PM, four days lags (average exposure over days $t,\dots,t-3$) for temperature, and no lag (average exposure over day $t$) for humidity.
As reference values, we use \SI{10}{\micro\gram/\metre^3} for \PM and \SI{15}{\celsius} for temperature.

The reference frames $\mathcal{T}(t)$ are constructed according to the time stratified design: each reference frame contains time-index corresponding to the same day of the week and within four weeks (28 days) of each others.
The vast majority of reference frames thus contain four observations (three control days); edge effects and the removal of holidays occasionally yield smaller reference frames.

\begin{figure}[t]
    \includegraphics[scale=1]{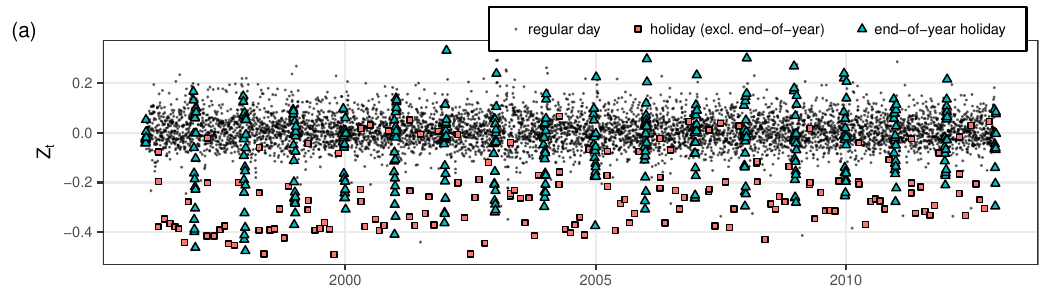}
    \includegraphics[scale=1]{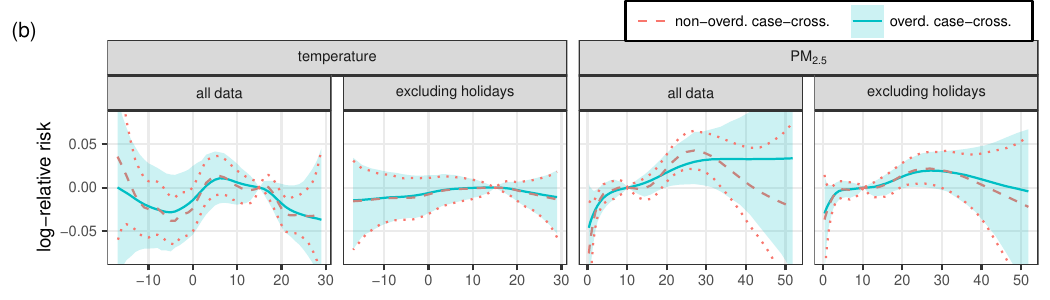}
    \caption{Results of the application presented in Section~\ref{sec:application}.
    Panel (a) shows the posterior medians of the overdispersion effects ($Z_t$) from the overdispersed case-crossover design (fitted using all data) and plotted against the corresponding date ($t$).
    Effects corresponding to the end-of-year holiday periods are identified by triangles, and the remaining public holidays in Toronto by squares.
    Panel (b) shows the median exposure-response curves and its 90\% global envelope for \PM and temperature, obtained by fitting the overdispersed (red) and non-overdipsersed (blue, dotted lines) case-crossover models to the two datasets considered in the application of Section~\ref{sec:application}. The credible intervals for the non-overdispersed model are shows as dotted lines.} \label{fig:app}
\end{figure}

\subsection{Analysis and results}

As a first step, we fitted the overdispersed case-crossover model to the full dataset (including holidays) and inspected the posterior distributions of the overdispersion effects $Z_t$.
Figure~\ref{fig:app}.a, which depicts their posterior medians, shows the presence of several outliers, mostly low ones.
It further shows that holidays indeed correspond, for the most part, to outliers, and in fact that they account for almost all the outliers in the data.
To assess the robustness of the overdispersed case-crossover model with respect to these frequently excluded observations, we re-fitted it on the data excluding holidays.
For comparison, we also fitted the standard (non-overdispersed) model to both datasets (including and excluding holidays).

Figure~\ref{fig:app}.b compares the posterior medians and 90\% global envelopes of the estimated log-relative risks of temperature and \PM on daily morbidity for the four combinations of models and datasets.
A notable feature is that the shaded regions showing posterior regions from the overdispersed model are much more consistent across the two datasets than the dotted lines from the standard non-overdispersed model.
The non-overdispersed model has an implausible protective effect of \PM for values above 30 when using the full dataset, whereas the overdispersed model shows an estimated effect consistent with results reported elsewhere\cite{Liu:2019}.  As expected, removing holidays causes the two models to become more similar, and the estimated standard deviation of the overdispersion $\sigma_0$ drops to $0.04$ from the $0.12$ estimated with the full dataset.  The non-overdispersed model produces a plausible exposure-response curve when holidays are excluded, although suggestion of a decreasing effect on the right is stronger than for the overdispersed model.  This result is consistent with the findings of the simulation study that the advantage of the overdispersed model is small, but still apparent, when the overdispersion is modest.

\section{Conclusions}\label{sec:conclusions}

In this paper, we demonstrated the compatibility and necessity of overdispersion within the case-crossover paradigm.
In the standard formulation, case-crossover models implicitly assume independence between subjects and, as a consequence, the absence of overdispersion.
Explicitly introducing shared overdispersion random effects into the individual-specific hazard functions gives a simple and intuitive framework for incorporating overdispersion in case-crossover models.
These time-varying effects, which are not assumed to be smoothly varying, do not cancel out in the partial likelihood \eqref{eq:likelihood} and hence they induce dependence in the responses for different individuals, which causes overdispersion in the population-level, aggregated model given in \eqref{eq:likelihood}.

An opinion that conditional logistic models are incompatible with overdispersion has been expressed previously.\cite{Lu/Zeger:2007, Armstrong/Gasparrini/Tobias:2014}
By characterizing overdispersion as residual dependence, we clarified the relationship between traditional case-crossover (i.e., conditional logistic) models, which operate at the individual level, and conditional Poisson models like that of Armstrong~et~al. (2014)\cite{Armstrong/Gasparrini/Tobias:2014}, which operate at the aggregated level.
In contrast with Armstrong~et~al. (2014), who rely on quasi-likelihood methods to account for overdispersion, our approach has the advantage of making the overdispersion an explicit component of the individual-level model.
However, Armstrong et al. (2014) further consider residual autocorrelation in addition to overdispersion, using the conditional quasi-likelihood of Brumback et al. (2000)\cite{Brumback/al:2000} for inference.
In our framework, this could be accomplished with an autoregressive model for the overdispersion effects; this more general model for the residual dependence could prove beneficial when larger reference frames are used.


As we demonstrated in the simulation study of Section~\ref{sec:sim-study}, failing to account for overdispersion when using the case-crossover designs can lead to dangerously underestimated confidence/credible intervals for the parameters of inferential interest.
In Section \ref{sec:application}, we stressed this point by highlighting, in a standard air pollution epidemiology analysis, the robustness of the overdispersed case-crossover model with respect to holiday-driven outlying observations.
We showed that identifying outliers, which can be done via the fitted values for the overdispersion coefficients, and removing them can significantly reduce the uncertainty around the estimates of the parameters of interest.
Thus, McCullagh and Nelder's statement that ``it seems wise to be cautious and to assume that over-dispersion is present to some extent unless and until it is shown to be absent''\cite{McCullagh/Nelder:1989} remains relevant, even when a partial likelihood is canceling out much of the data's heterogeneity.

\section*{Acknowledgments}

Samuel Perreault was funded by the Institute de valorisation de donn\'{e}es (IVADO) and the Fonds de Recherche du Qu\'{e}bec -- Nature et technologies (FRQNT).
Gracia Dong was funded by the Canadian Statistical Sciences Institute (CANSSI).
Alex Stringer was funded by the National Sciences and Engineering Research Council of Canada (NSERC, RGPIN-2023-03331).
Patrick Brown was funded by the National Sciences and Engineering Research Council of Canada (NSERC, RGPIN-2022-05164).
Data for this study was provided by Statistics Canada (mortality in vital statistics), the Canadian Institute for Health Information (hospitalizations) and Environment and Climate Change Canada (air pollution and weather).
We gratefully acknowledge data stewardship and contribution of all data providers.
However, the analyses, conclusions, opinions and statements expressed herein are not necessarily those of the data providers.
We also gratefully acknowledge the help of Guowen Huang and Hana Fu in preparing the data for analysis.

\section*{Supporting information}

\textbf{Figures~\ref{fig:time-trend}--\ref{fig:E2-ci-p}}, which provides a detailed summary of the results for the simulation study of Section~\ref{sec:sim-study}, are available as part of the web appendix.



\bibliographystyle{vancouver}


\clearpage

\appendix
\renewcommand{\thesection}{\Alph{section}}
\setcounter{section}{1}
\setcounter{figure}{0}
\renewcommand\thefigure{\thesection.\arabic{figure}} 
\section*{Appendix}

\begin{figure}[h]
\centering
\includegraphics[scale=1]{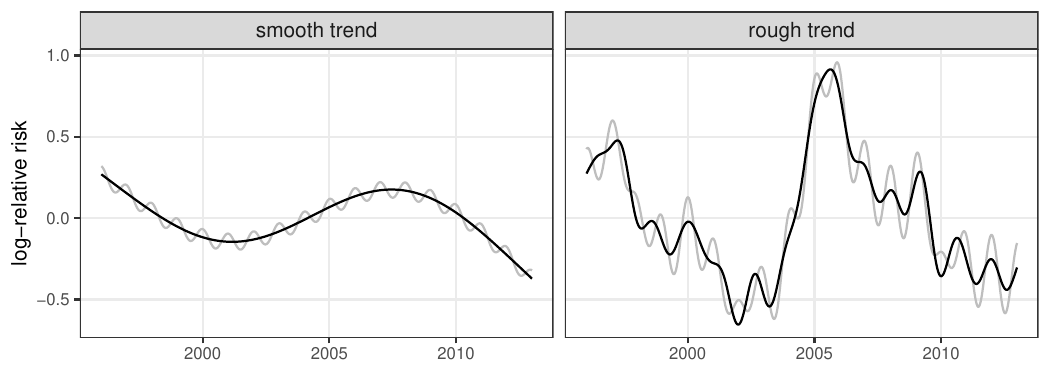}
\caption{Depiction of the two time trends used in the simulation experiments of Section~3. The black curves show the (log) magnitude of the long-term trend, while the gray curves show the combined (log) magnitude of the seasonal and long-term trends.} \label{fig:time-trend}
\end{figure}

\begin{figure}[p]
\centering
\includegraphics[scale=1]{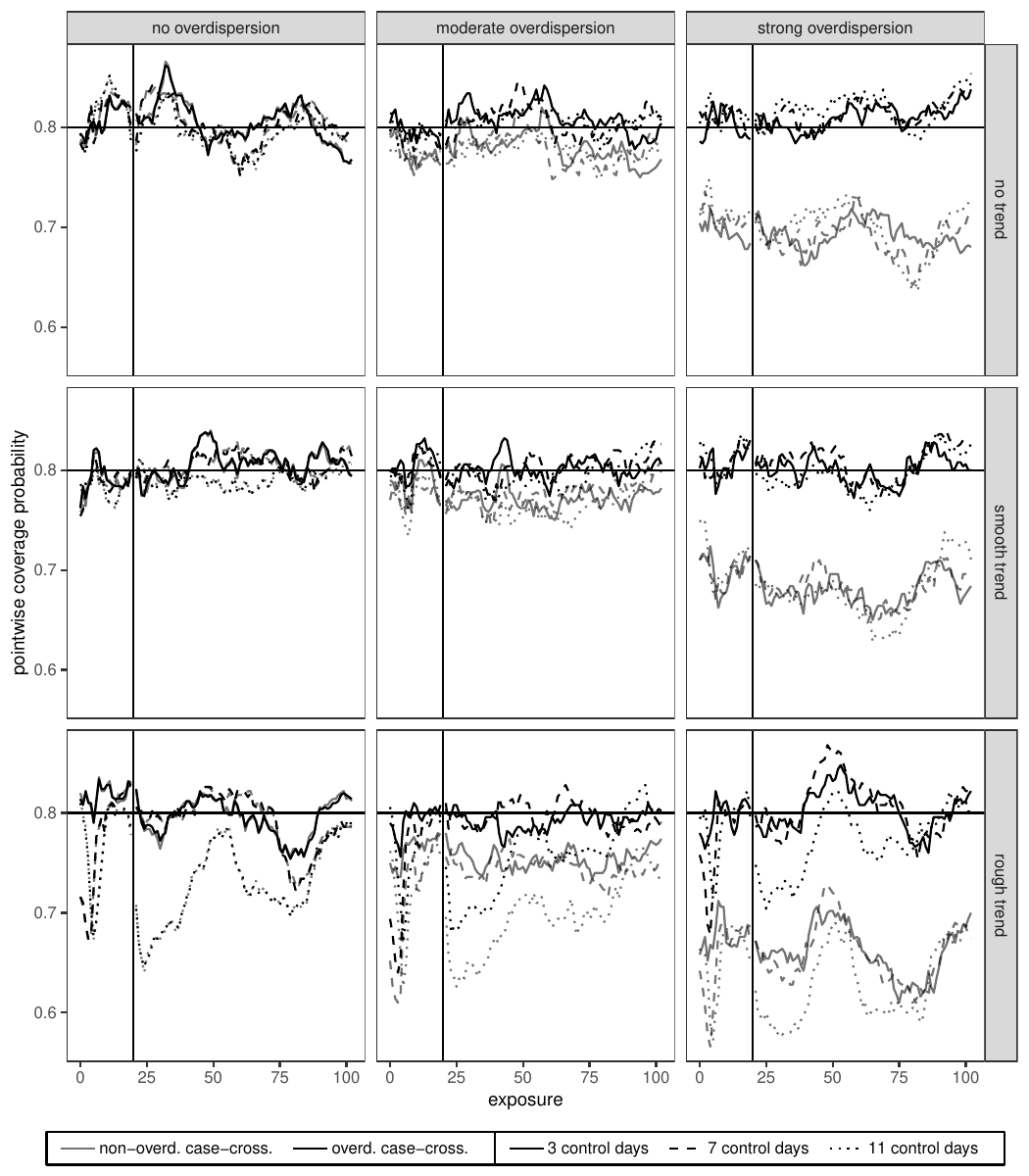}
\caption{Results from Experiment I (Section 3.2): Estimated coverage probabilities of the pointwise 80\% credible intervals from the non-overdispersed and overdispersed case-crossover models (time stratified with 3, 7 and 11 control days), for each time trend and level of overdispersion.
The horizontal and vertical lines correspond to 80\% and the reference value 20, respectively.
The results are based on 500 replications.
} \label{fig:E1-pc}
\end{figure}

\begin{figure}[p]
\centering
\includegraphics[scale=1]{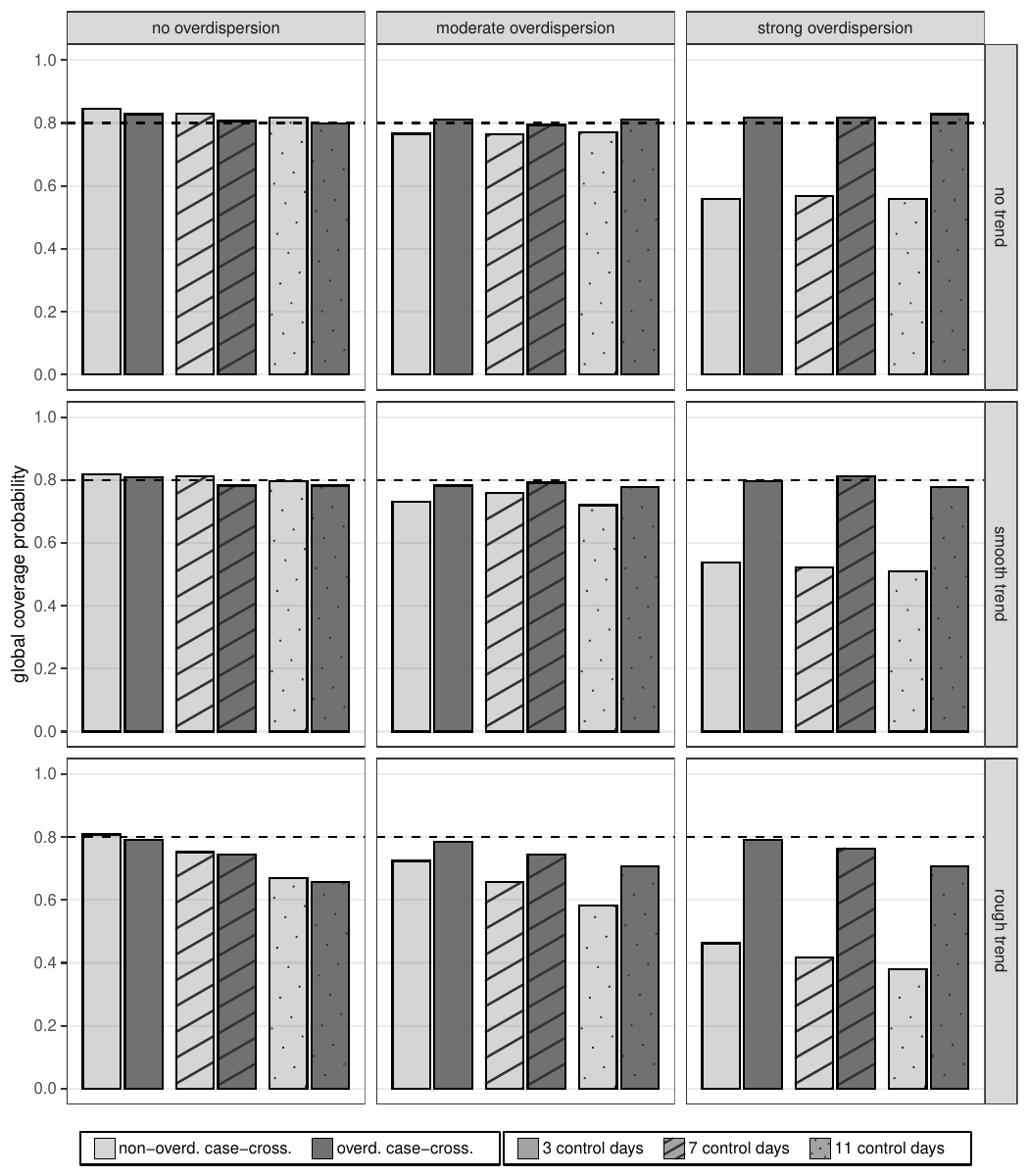}
\caption{Results from Experiment I (Section 3.2): Estimated coverage probabilities of the 80\% global envelopes from the non-overdispersed and overdispersed case-crossover models (time stratified with 3, 7 and 11 control days), for each time trend and level of overdispersion.
The horizontal lines correspond to 80\%.
The results are based on 500 replications.
} \label{fig:E1-ge}
\end{figure}

\begin{figure}[p]
\centering
\includegraphics[scale=1]{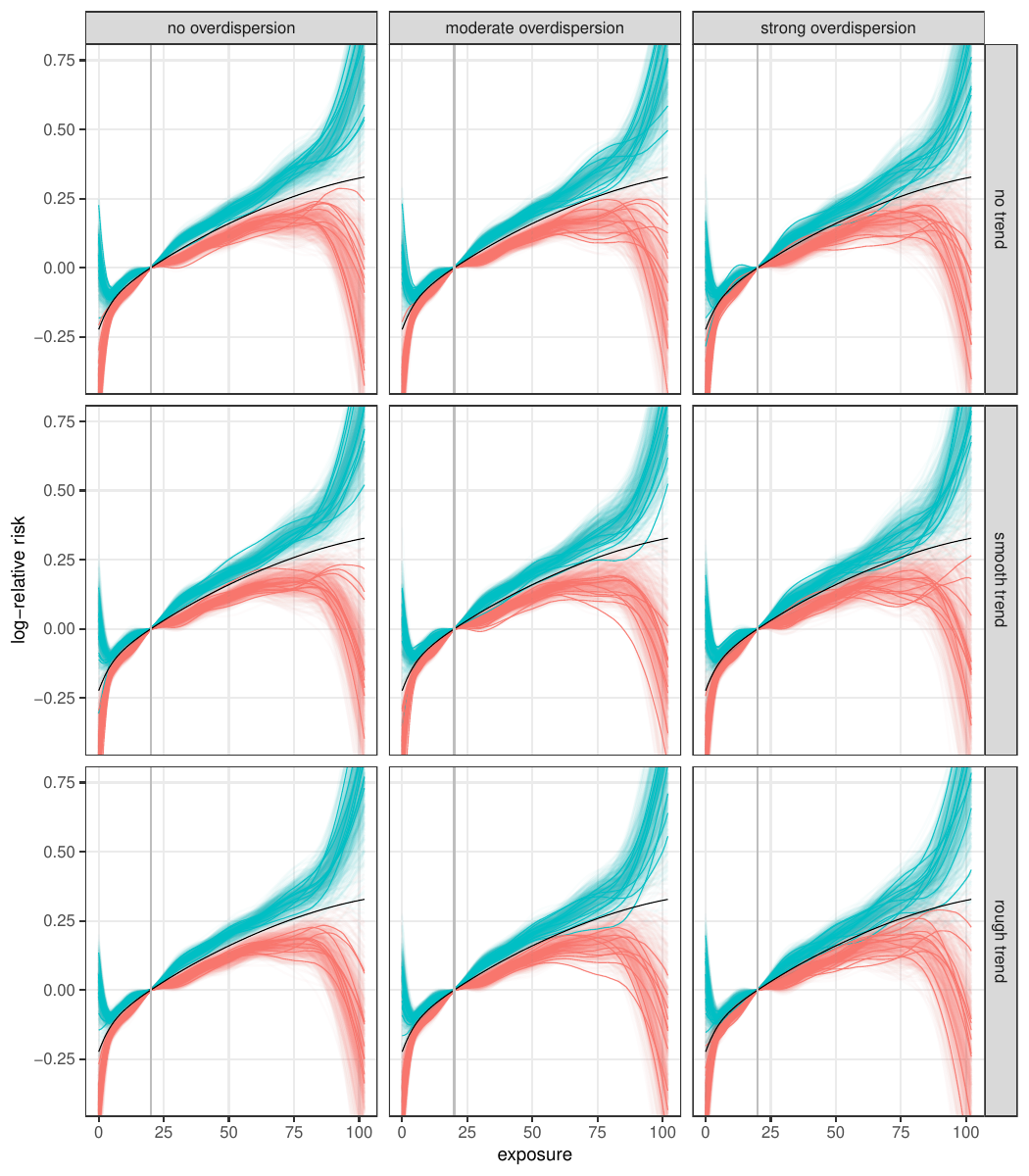}
\caption{Results from Experiment I (Section 3.2): The 500 estimated $80\%$ global envelopes returned by the \textbf{non-overdispersed} case-crossover model (time stratified, \textbf{three control days}), with ten randomly selected envelopes shown more prominently.
The black curve depicts the true log-relative risk and the dark gray vertical line corresponds to the reference value 20.} \label{fig:E1-ci-nod3}
\end{figure}

\begin{figure}[p]
\centering
\includegraphics[scale=1]{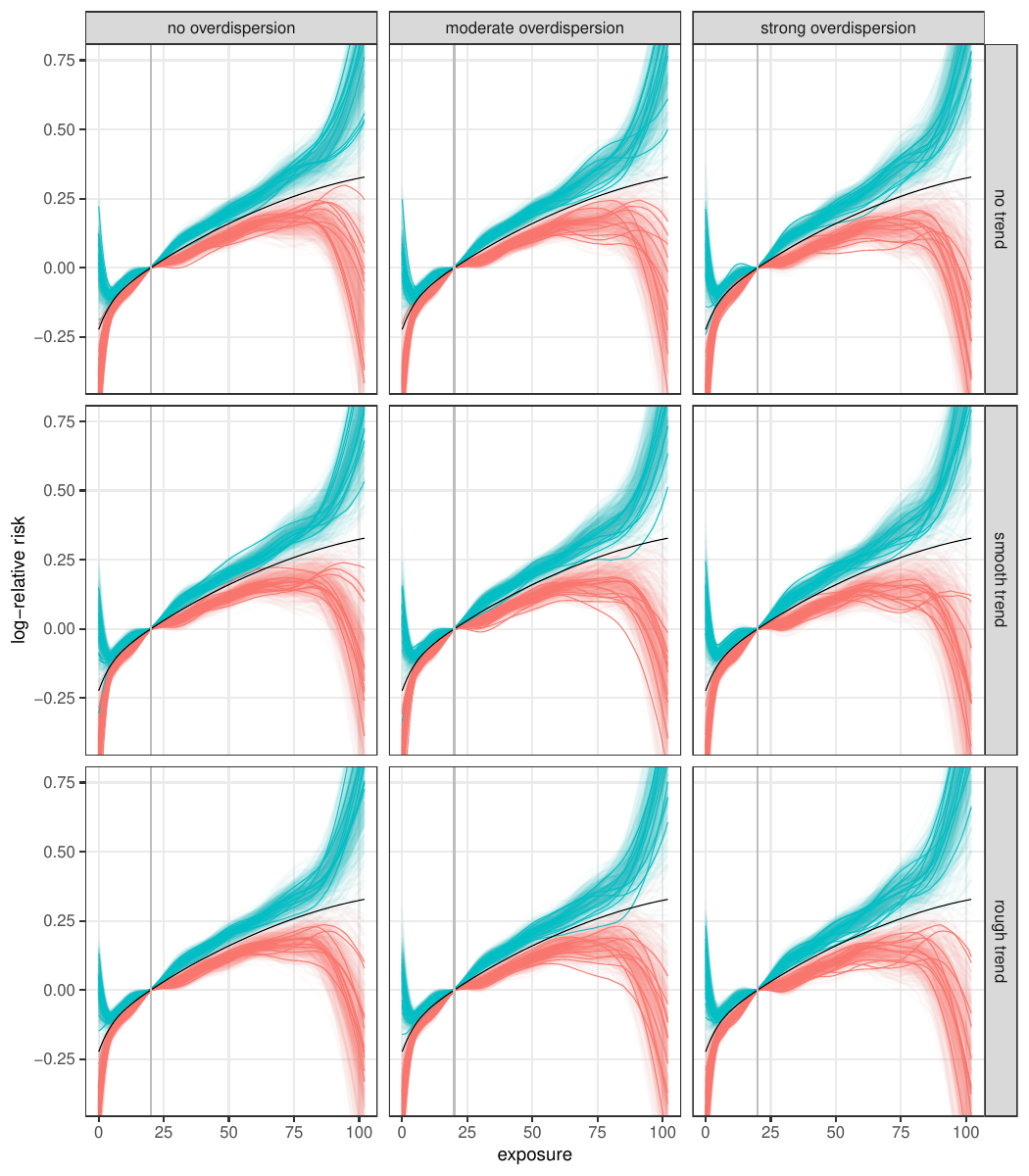}
\caption{Results from Experiment I (Section 3.2): The 500 estimated $80\%$ global envelopes returned by the \textbf{overdispersed} case-crossover model (time stratified, \textbf{three control days}), with ten randomly selected envelopes shown more prominently.
The black curve depicts the true log-relative risk and the dark gray vertical line corresponds to the reference value 20.} \label{fig:E1-ci-od3}
\end{figure}

\begin{figure}[p]
\centering
\includegraphics[scale=1]{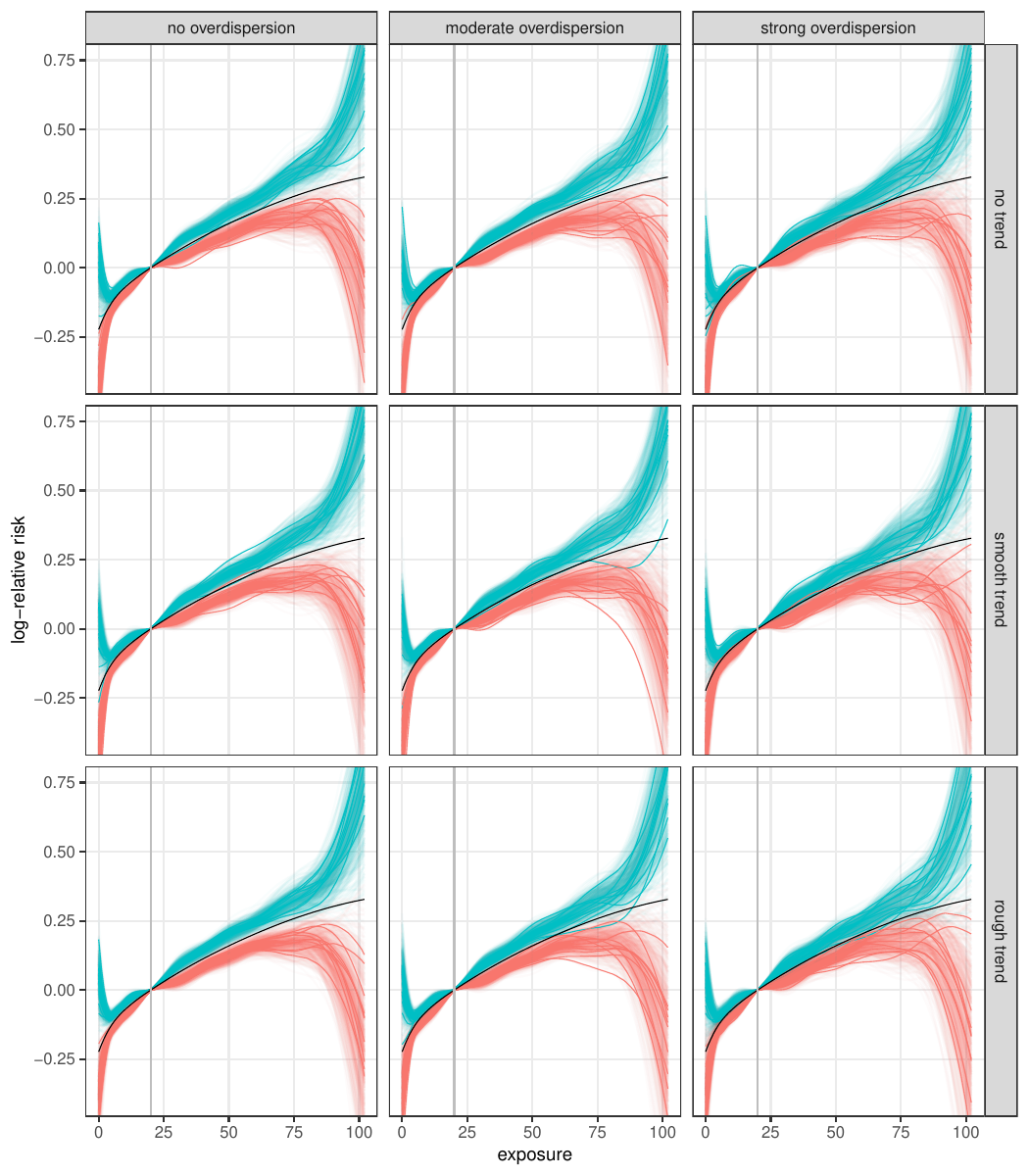}
\caption{Results from Experiment I (Section 3.2): The 500 estimated $80\%$ global envelopes returned by the \textbf{non-overdispersed} case-crossover model (time stratified, \textbf{seven control days}), with ten randomly selected envelopes shown more prominently.
The black curve depicts the true log-relative risk and the dark gray vertical line corresponds to the reference value 20.} \label{fig:E1-ci-nod7}
\end{figure}

\begin{figure}[p]
\centering
\includegraphics[scale=1]{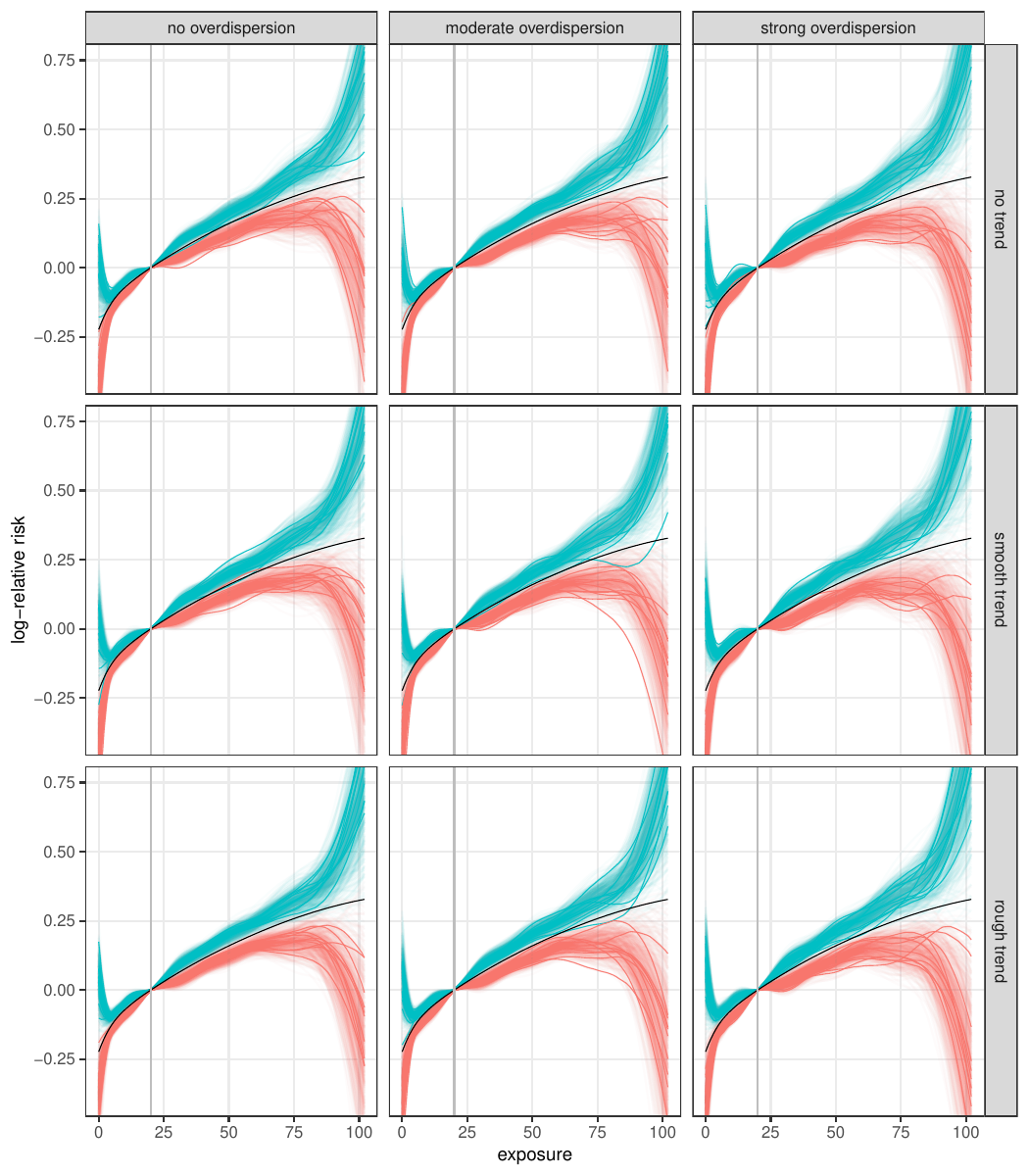}
\caption{Results from Experiment I (Section 3.2): The 500 estimated $80\%$ global envelopes returned by the \textbf{overdispersed} case-crossover model (time stratified, \textbf{seven control days}), with ten randomly selected envelopes shown more prominently.
The black curve depicts the true log-relative risk and the dark gray vertical line corresponds to the reference value 20.} \label{fig:E1-ci-od7}
\end{figure}

\begin{figure}[p]
\centering
\includegraphics[scale=1]{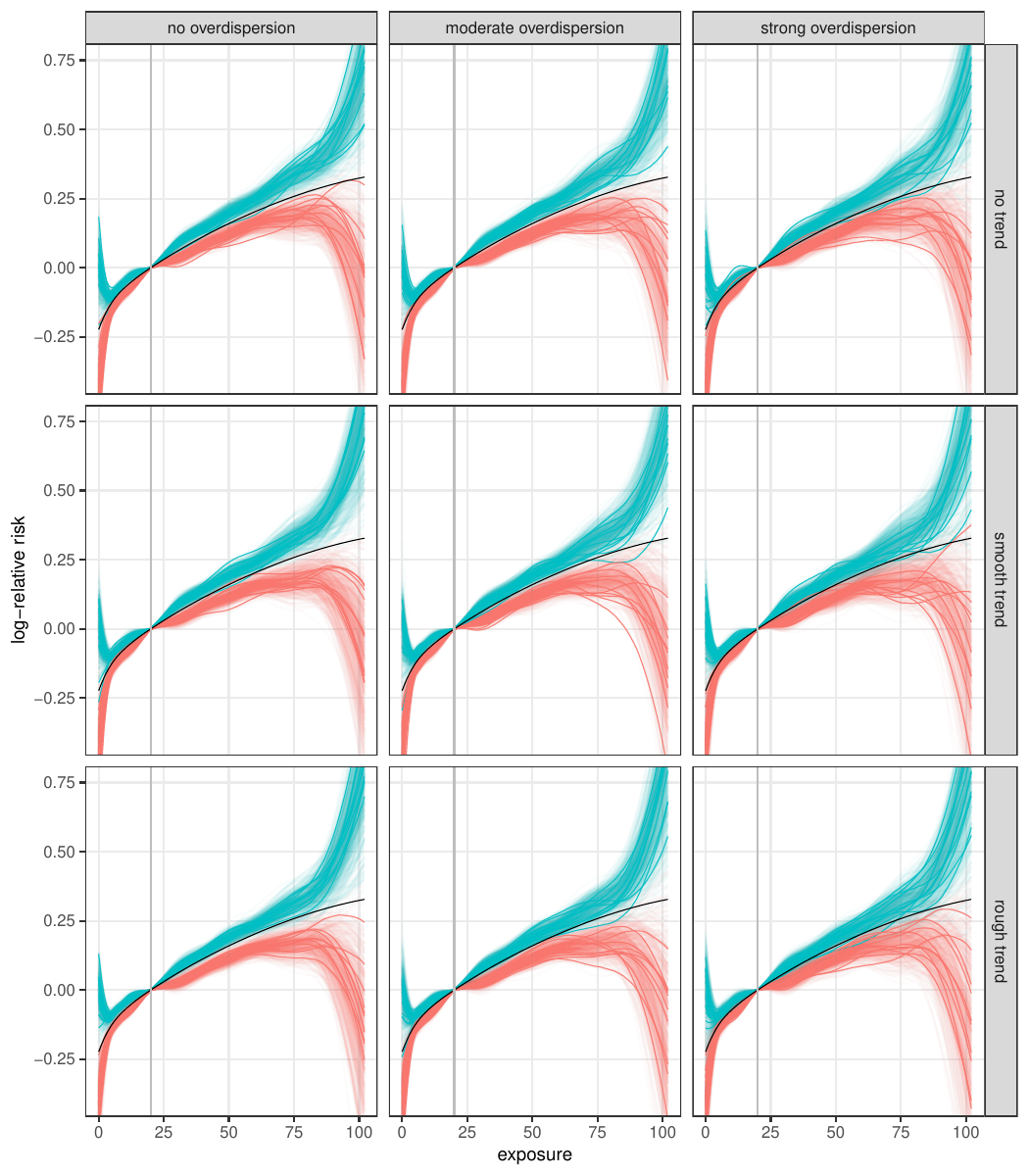}
\caption{Results from Experiment I (Section 3.2): The 500 estimated $80\%$ global envelopes returned by the \textbf{non-overdispersed} case-crossover model (time stratified, \textbf{eleven control days}), with ten randomly selected envelopes shown more prominently.
The black curve depicts the true log-relative risk and the dark gray vertical line corresponds to the reference value 20.} \label{fig:E1-ci-nod11}
\end{figure}

\begin{figure}[p]
\centering
\includegraphics[scale=1]{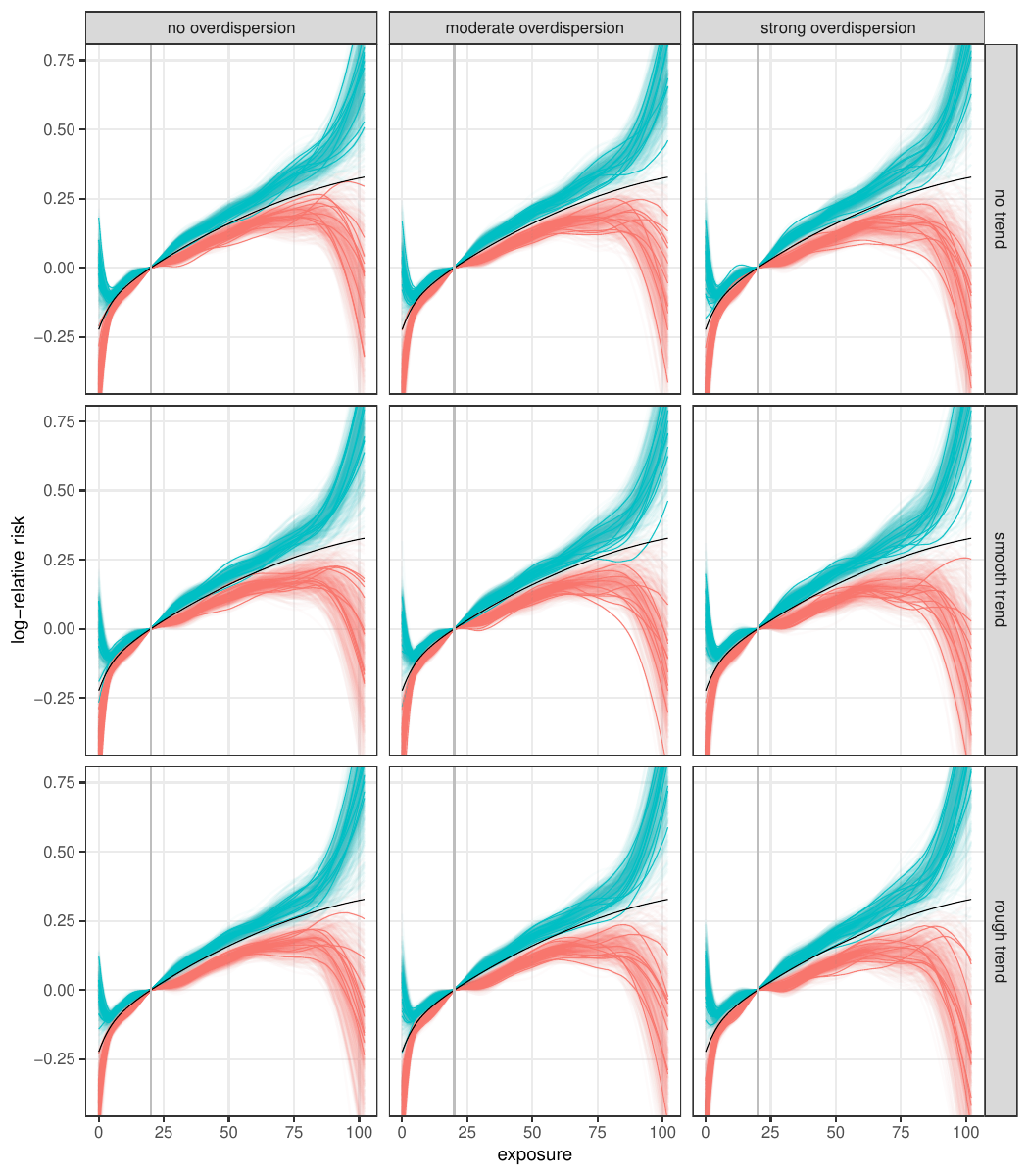}
\caption{Results from Experiment I (Section 3.2): The 500 estimated $80\%$ global envelopes returned by the \textbf{overdispersed} case-crossover model (time stratified, \textbf{eleven control days}), with ten randomly selected envelopes shown more prominently.
The black curve depicts the true log-relative risk and the dark gray vertical line corresponds to the reference value 20.} \label{fig:E1-ci-od11}
\end{figure}

\begin{figure}[p]
\centering
\includegraphics[scale=1]{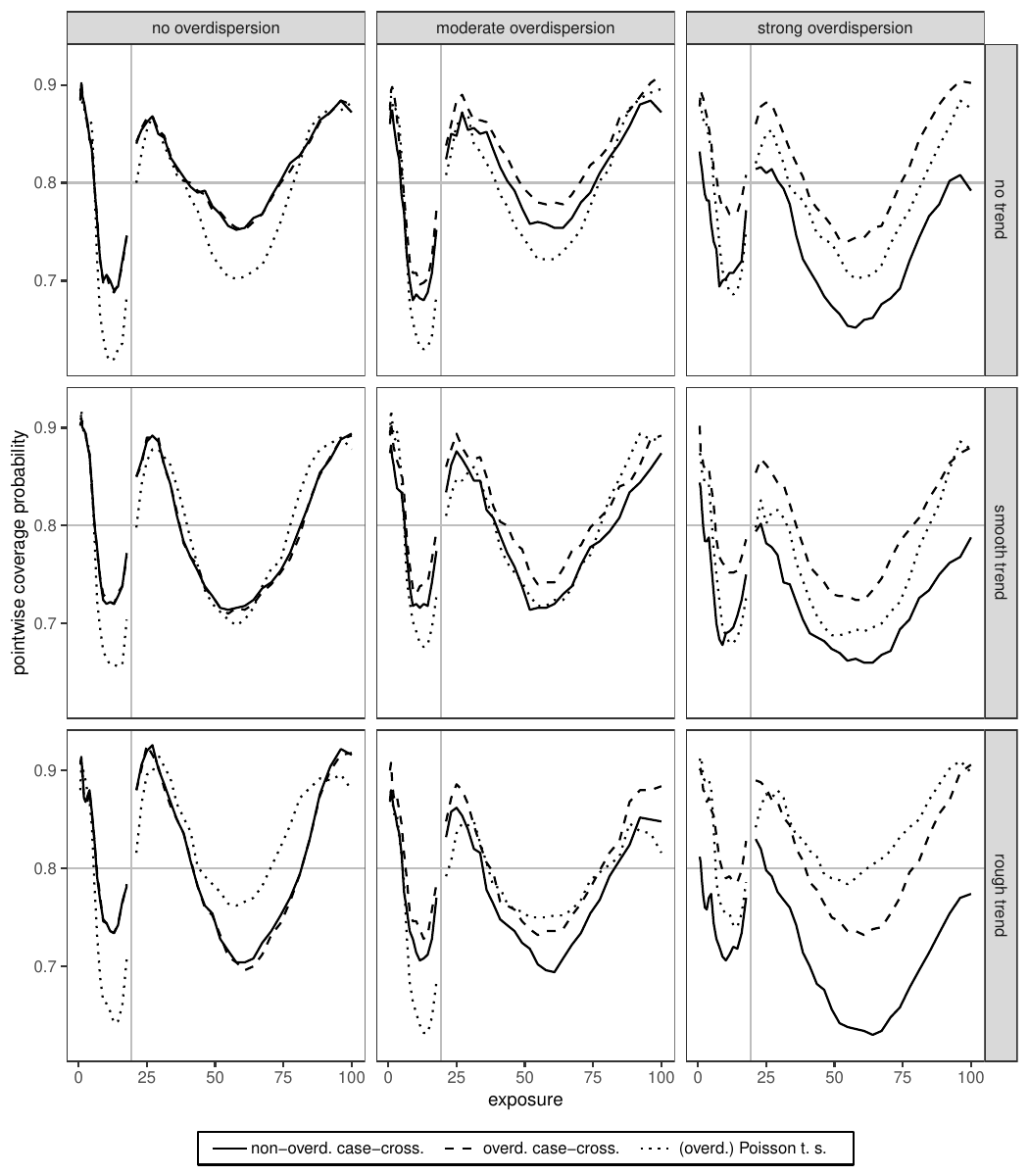}
\caption{Results from Experiment II (Section 3.3): Estimated coverage probabilities of the pointwise 80\% credible intervals from the non-overdispersed and overdispersed case-crossover models (time stratified, three control days) and the Poisson time series model, for each time trend and level of overdispersion.
The horizontal and vertical lines correspond to 80\% and the reference value 20, respectively.
The results are based on 500 replications.} \label{fig:E2-pc}
\end{figure}

\begin{figure}[p]
\centering
\includegraphics[scale=1]{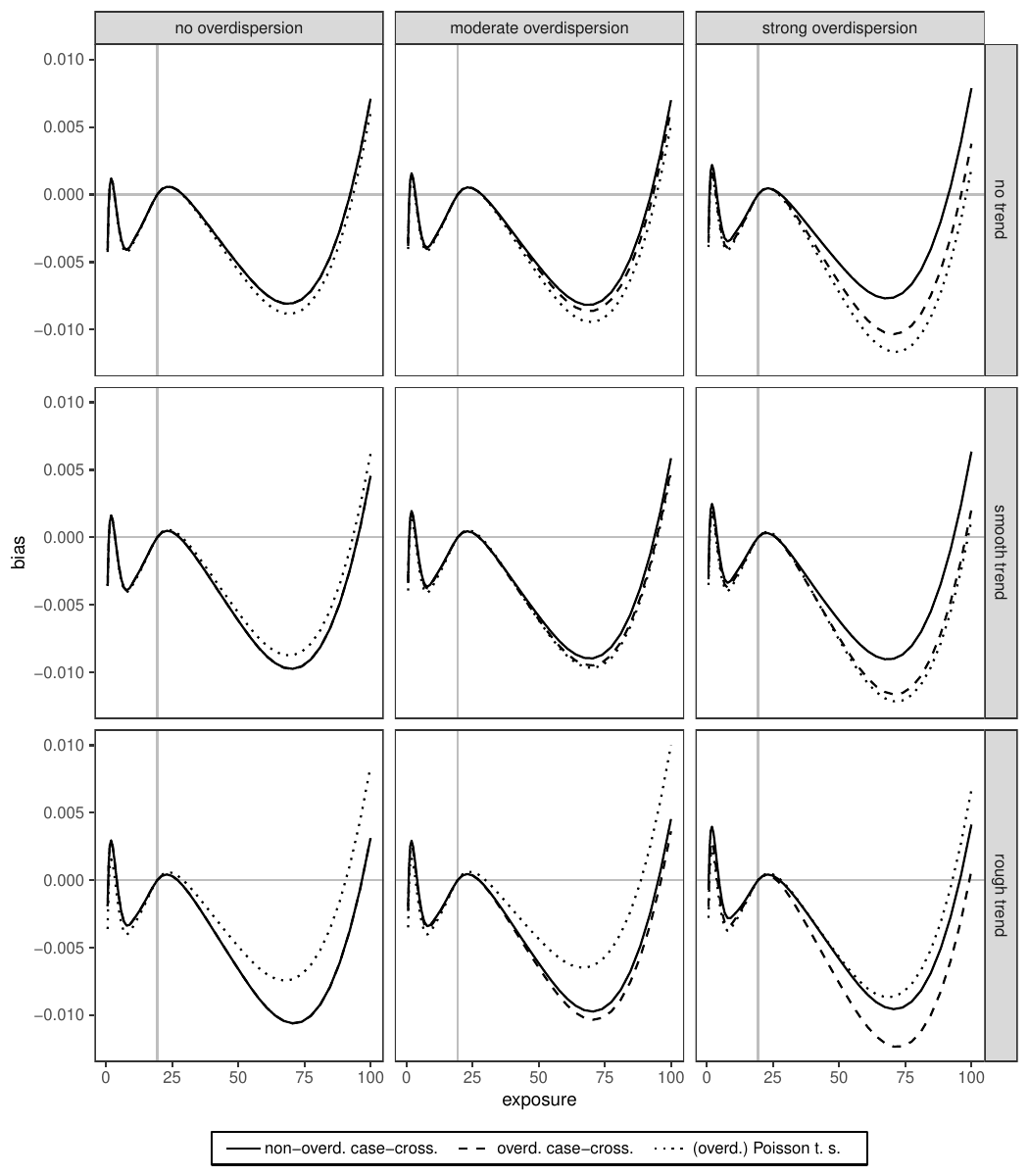}
\caption{Results from Experiment II (Section 3.3): Estimated bias of the posterior median point estimates from the non-overdispersed and overdispersed case-crossover models (time stratified, three control days) and the Poisson time series model, for each time trend and level of overdispersion.
The gray horizontal and vertical lines correspond to 0 (no bias) and the reference value 20, respectively
The results are based on 500 replications.} \label{fig:E2-bias}
\end{figure}

\begin{figure}[p]
\centering
\includegraphics[scale=1]{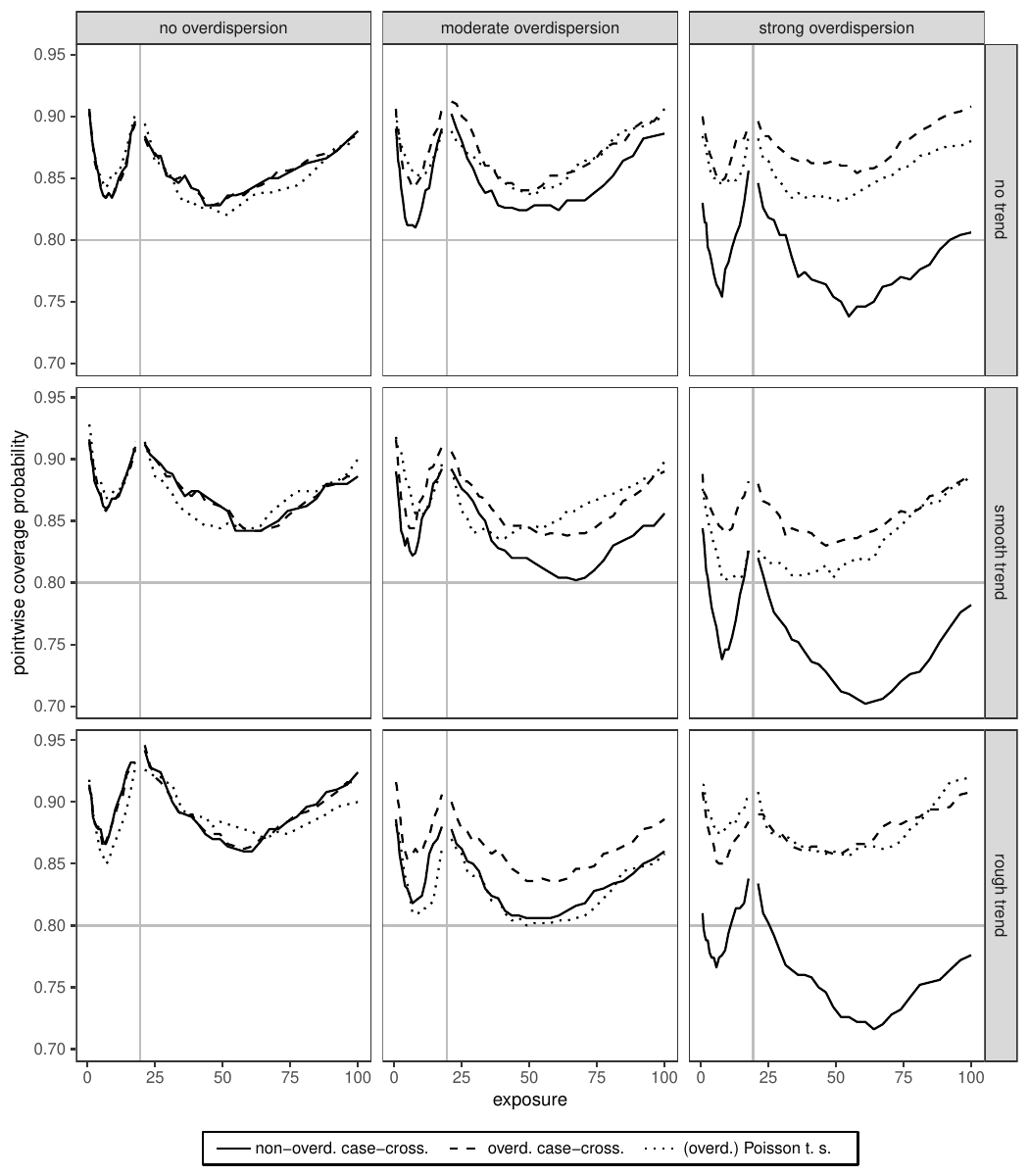}
\caption{Results from Experiment II (Section 3.3): Estimated coverage probabilities of the \textbf{bias-adjusted} pointwise 80\% credible intervals from the non-overdispersed and overdispersed case-crossover models (time stratified, three control days) and the Poisson time series model, for each time trend and level of overdispersion.
These are obtained by subtracting the bias shown in Figure~\ref{fig:E2-bias} to the original pointwise 80\% credible intervals.
The horizontal and vertical lines correspond to 80\% and the reference value 20, respectively.
The results are based on 500 replications.} \label{fig:E2-pc-adj}
\end{figure}

\begin{figure}[p]
\centering
\includegraphics[scale=1]{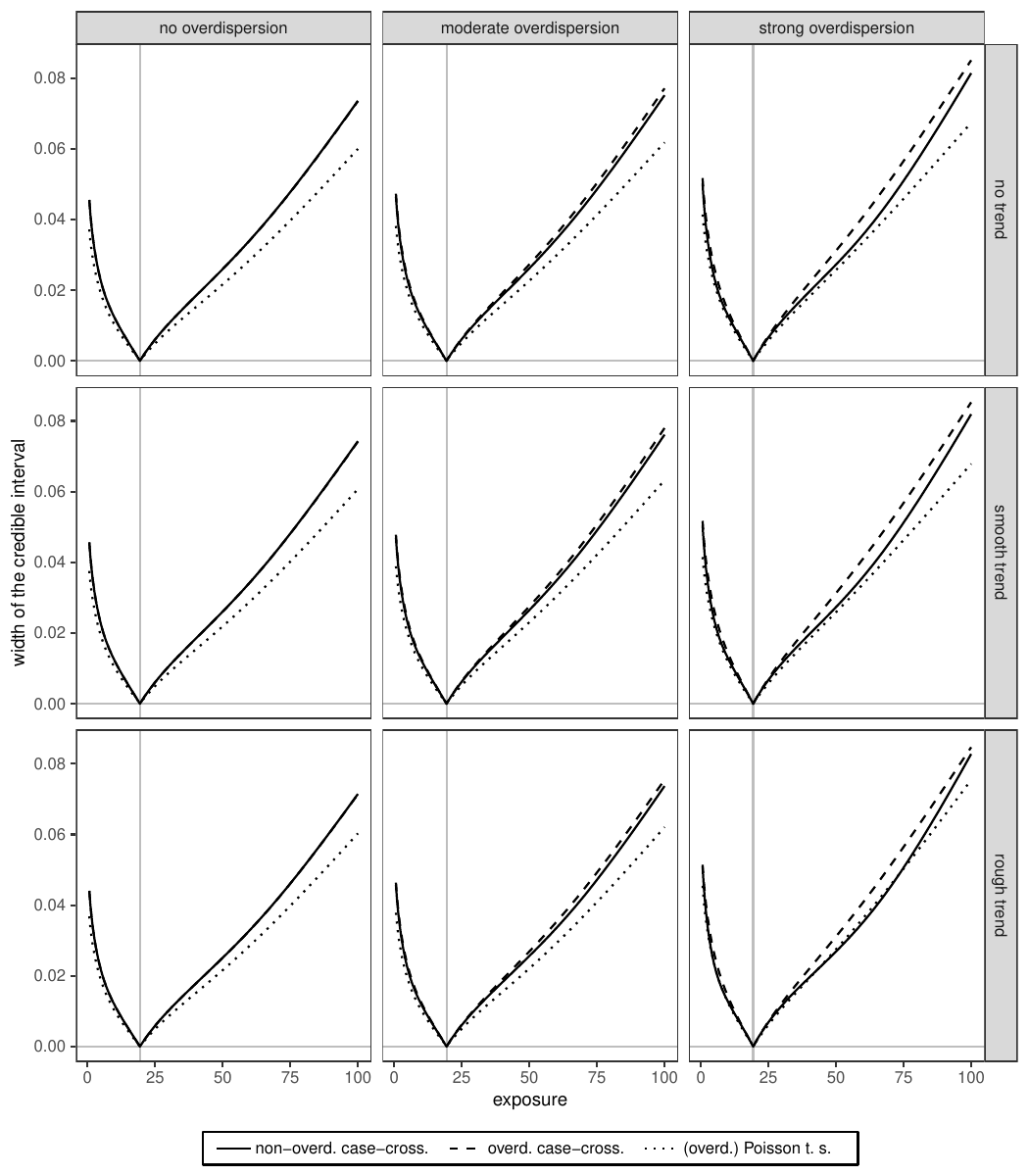}
\caption{Results from Experiment II (Section 3.3): Estimated average width of the pointwise 80\% credible intervals from the non-overdispersed and overdispersed case-crossover models (time stratified, three control days) and the Poisson time series model, for each time trend and level of overdispersion.
The gray vertical lines correspond to the reference value 20.
The results are based on 500 replications.} \label{fig:E2-width}
\end{figure}

\begin{figure}[p]
\centering
\includegraphics[scale=1]{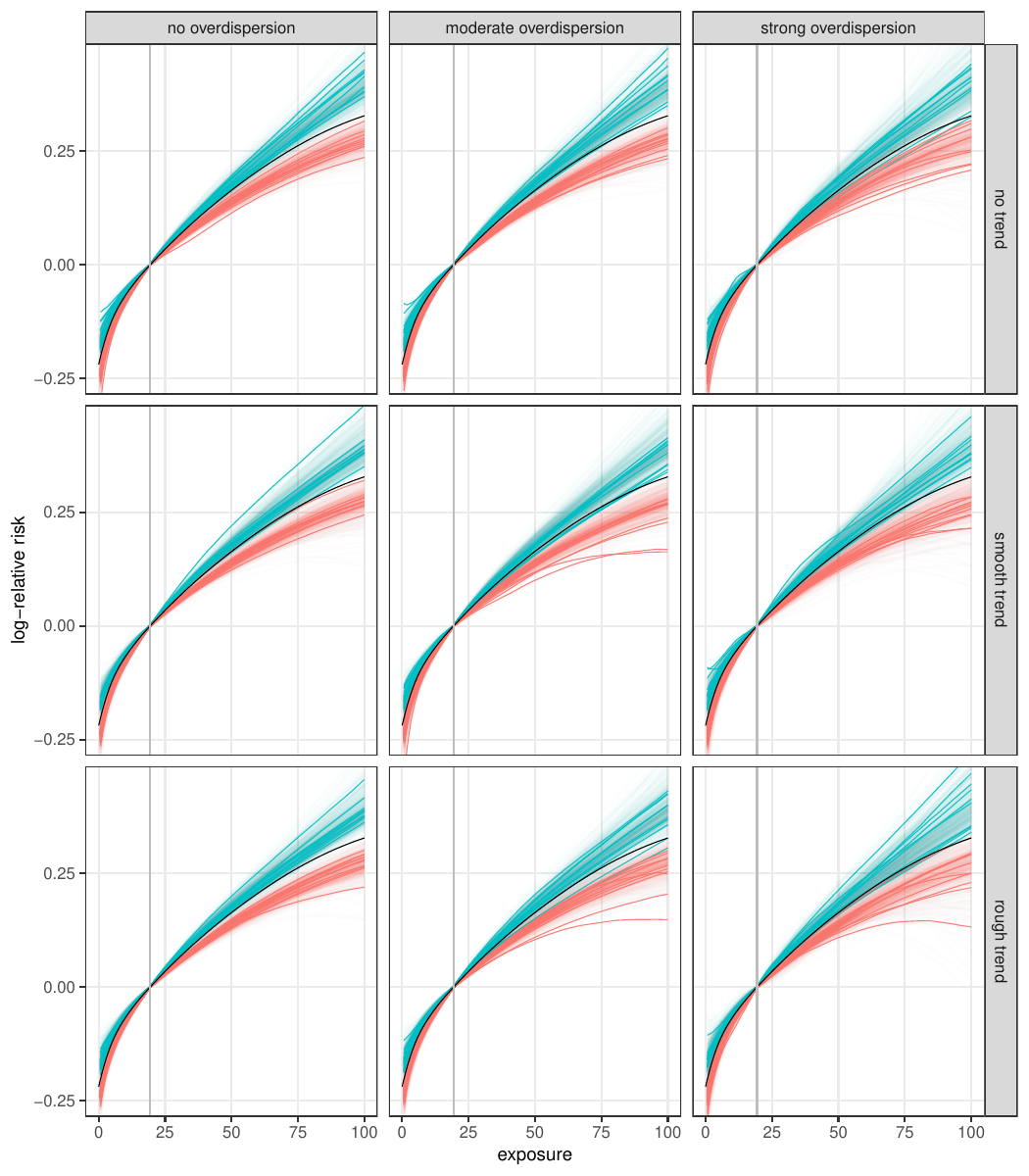}
\caption{Results from Experiment II (Section 3.3): The 500 estimated $80\%$ global envelopes returned by the \textbf{non-overdispersed} case-crossover model (time stratified, three control days), with ten randomly selected envelopes shown more prominently.
The black curve depicts the true log-relative risk and the dark gray vertical line corresponds to the reference value 20.} \label{fig:E2-ci-nod}
\end{figure}

\begin{figure}[p]
\centering
\includegraphics[scale=1]{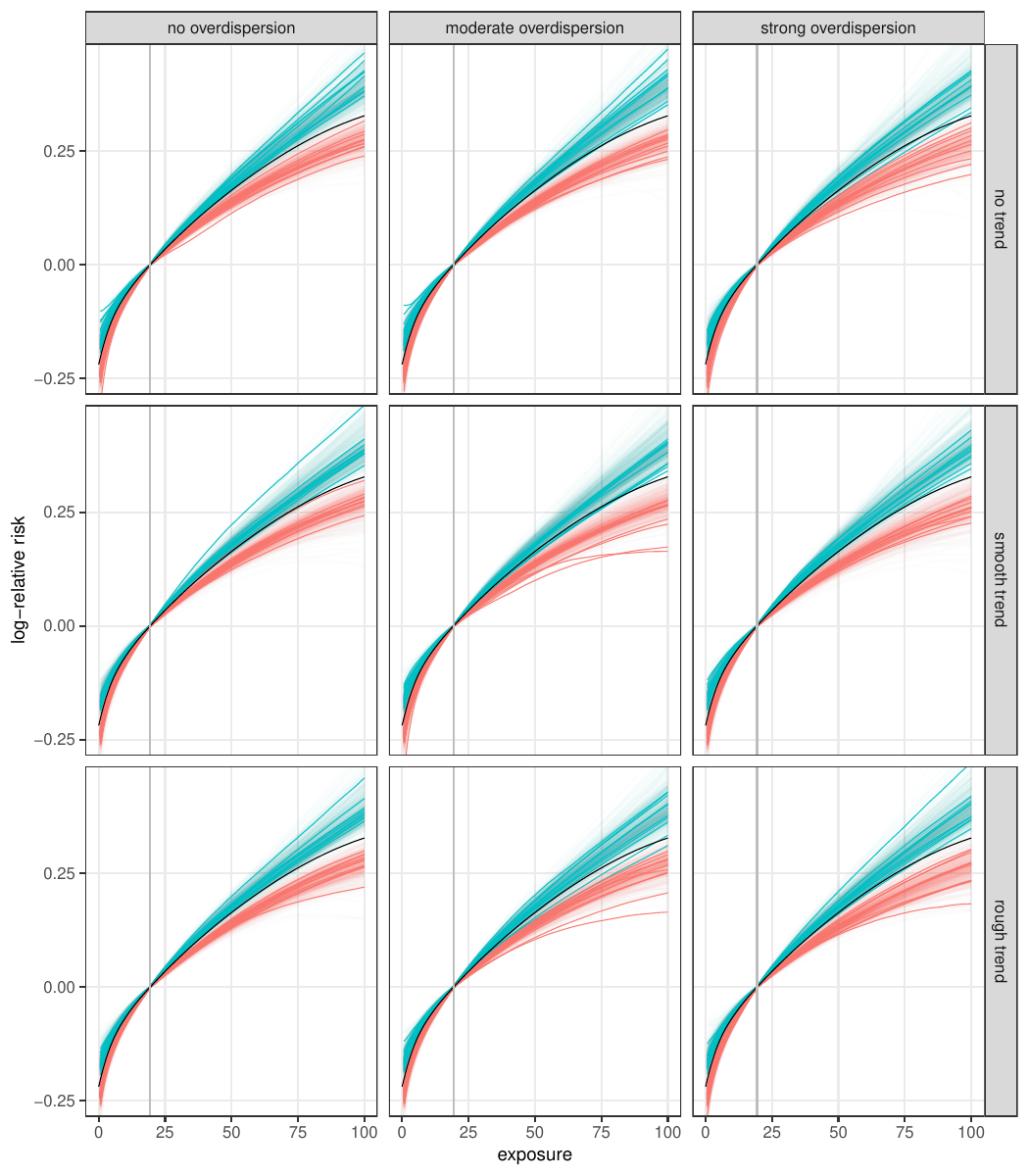}
\caption{Results from Experiment II (Section 3.3): The 500 estimated $80\%$ global envelopes returned by the \textbf{overdispersed} case-crossover model (time stratified, three control days), with ten randomly selected envelopes shown more prominently.
The black curve depicts the true log-relative risk and the dark gray vertical line corresponds to the reference value 20.} \label{fig:E2-ci-od}
\end{figure}

\begin{figure}[p]
\centering
\includegraphics[scale=1]{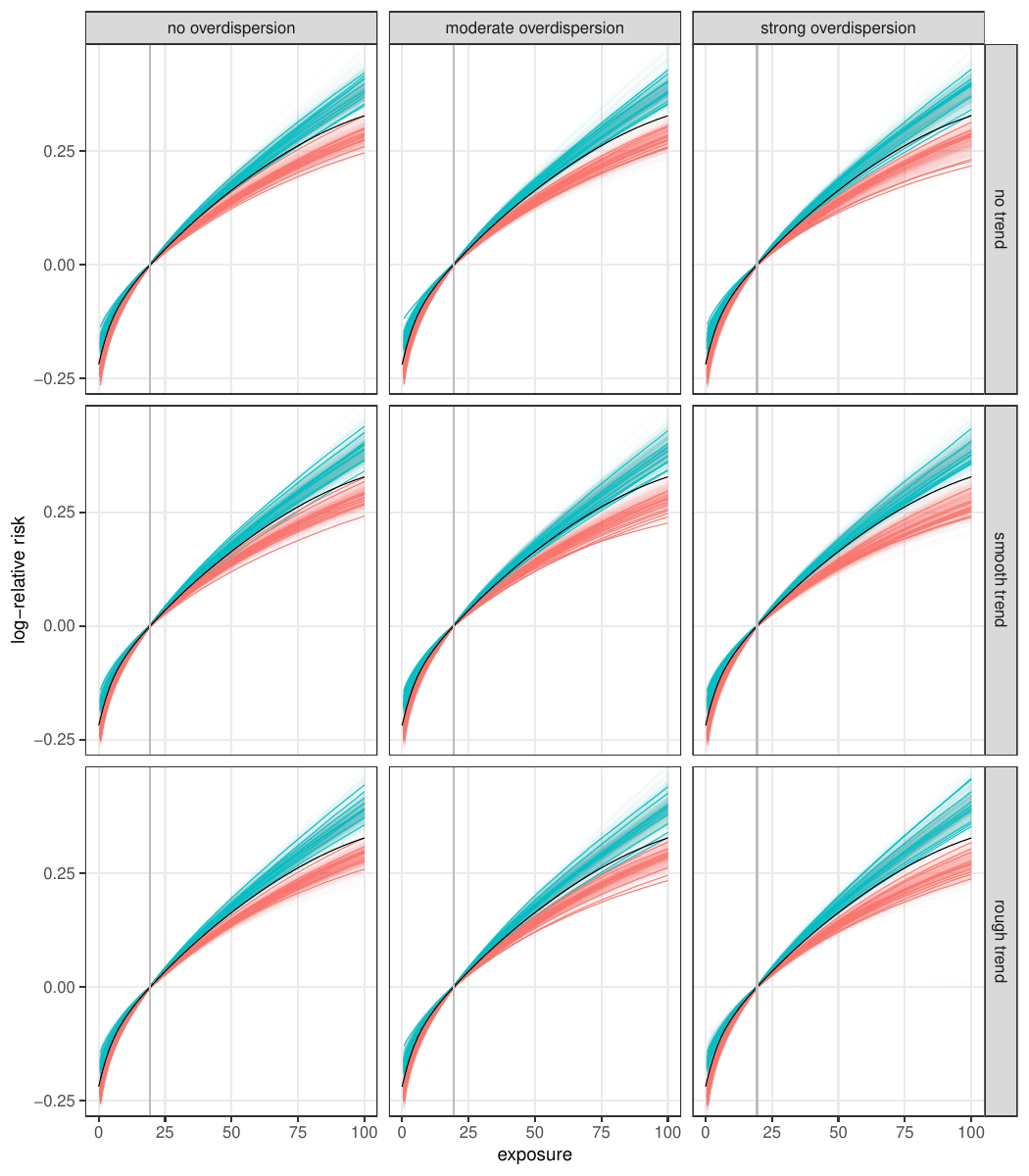}
\caption{Results from Experiment II (Section 3.3): The 500 estimated $80\%$ global envelopes returned by the \textbf{Poisson} time series model, with ten randomly selected envelopes shown more prominently.
The black curve depicts the true log-relative risk and the dark gray vertical line corresponds to the reference value 20.} \label{fig:E2-ci-p}
\end{figure}

\normalsize
\noindent Figure~\ref{fig:time-trend}, just below, depicts the time trends used to generate data for the simulation study of Section~\ref{sec:sim-study}. The remaining figures provide a comprehensive summary of the simulation results:
\begin{itemize}
    \item Figures~\ref{fig:E1-pc}--\ref{fig:E1-ci-od11} summarize Section~3.2, 
    which involves two case-crossover models (non-overdispersed and overdipsersed), each fitted using three designs (time stratified with 3, 7 and 11 control days).
    \begin{itemize}
        \item Figures~\ref{fig:E1-pc} depicts the coverage of $80\%$ pointwise credible intervals for all models and designs.
        \item Figures~\ref{fig:E1-ge} depicts the coverage of $80\%$ global envelops for all models and designs.
        \item Figures~\ref{fig:E1-ci-nod3}--\ref{fig:E1-ci-od11} each show the actual $80\%$ global envelops returned by a specific combination of model and design.
    \end{itemize}
    \item Figures~\ref{fig:E2-pc}--\ref{fig:E2-ci-p} summarize Section~3.3, 
    which involves two case-crossover models (non-overdispersed and overdipsersed), each fitted using a time stratified design with 3 control days, as well as a Poisson time series model.
    \begin{itemize}
        \item Figures~\ref{fig:E2-pc} depicts the coverage of $80\%$ pointwise credible intervals for all models.
        \item Figures~\ref{fig:E2-bias} depicts the bias (pointwise) for all models.
        \item Figures~\ref{fig:E2-pc-adj} depicts the coverage of bias-adjusted $80\%$ pointwise credible intervals for all models.
        \item Figures~\ref{fig:E2-width} depicts the width of the $80\%$ pointwise credible intervals for all models.
        \item Figures~\ref{fig:E2-ci-nod}--\ref{fig:E2-ci-p} each show the actual $80\%$ global envelops returned by a specific model.
    \end{itemize}
\end{itemize}

\end{document}